\documentclass{article}

\usepackage{arxiv}

\usepackage[utf8]{inputenc} 
\usepackage[T1]{fontenc}    
\usepackage{hyperref}       
\usepackage{url}            
\usepackage{booktabs}       
\usepackage{amsfonts}       
\usepackage{amsmath}
\usepackage{nicefrac}       
\usepackage{microtype}      
\usepackage{lipsum}
\usepackage{graphicx}

\DeclareMathOperator*{\E}{\mathbb{E}}

\title{Digital phase-only holography using deep conditional generative models}

\author{
  Jannes~Gladrow \\
  Cavendish Laboratory \\
    University of Cambridge\\
    JJ Thomson Ave,
    Cambridge, UK\\
  \texttt{jg754@cam.ac.uk} \\
}

\begin{document}
\maketitle

\begin{abstract}
Holographic wave-shaping has found numerous applications across the physical sciences, especially since the development of digital spatial-light modulators (SLMs). A key challenge in digital holography consists in finding optimal hologram patterns which transform the incoming laser beam into desired shapes in a conjugate optical plane. The existing repertoire of approaches to solve this inverse problem is built on iterative phase-retrieval algorithms, which do not take optical aberrations and deviations from theoretical models into account. 
Here, we adopt a physics-free, data-driven, and probabilistic approach to the problem. Using deep conditional generative models such as Generative-Adversarial Networks (cGAN) or Variational Autoencoder (cVAE), we approximate conditional distributions of holograms for a given target laser intensity pattern.
In order to reduce the cardinality of the problem, we train our models on a proxy mapping relating an $8\times 8$-matrix of complex-valued spatial-frequency coefficients to the ensuing $100\times100$-shaped intensity distribution recorded on a camera. We discuss the degree of 'ill-posedness' that remains in this reduced problem and compare different generative model architectures in terms of their ability to find holograms that reconstruct given intensity patterns. Finally, we challenge our models to generalise to synthetic target intensities, where the existence of matching holograms cannot be guaranteed. We devise a forward-interpolating training scheme aimed at providing models the ability to interpolate in laser intensity space, rather than hologram space and show that this indeed enhances model performance on synthetic data sets.
\end{abstract}

\section{Introduction}
\label{sec:introduction}

With the advent of electrically addressed spatial light modulators (SLMs), digital holography has become a widely applied and researched technique~\cite{Dudley2003}. Each pixel in a SLM is able to independently modulate phase or amplitude of incoming light wavefronts. In particular, phase-only SLMs, which do not absorb any of the incoming light, have found various applications in experimental physics~\cite{Grier2003, Padgett2011, Curtis2012, Gladrow2019}, (holographic) data storage~\cite{Dhar2008, Zhang2014}, light detection and ranging (LiDAR) systems~\cite{Smith2017}, while research into enhancing their performance is ongoing~\cite{Yang2014, Smolyaninov2019, Yang2019}. However, improvements in spatial resolution and pixel refresh rates emphasise the need for progress on the algorithmic side tasked with computing suitable phase-shifting patterns. The algorithmic problem encountered in digital holography can be summarised as follows: find the hologram pattern, that, when displayed on the SLM, creates the closest approximation to a given desired intensity distribution in a conjugate optical plane.

This inverse problem is an instance of the phase-retrieval problem and is therefore severely complicated by phase-invariances of light intensity. The difficulty and importance of this problem have inspired decades of research into phase-retrieval algorithms and general approaches to inverse problems in imaging~\cite{Gerchberg1972, Boccacci1998, miao1998phase, DiLeonardo2007, Rivenson2018}. 
\begin{figure}
  \centering
    \includegraphics[width=154mm]{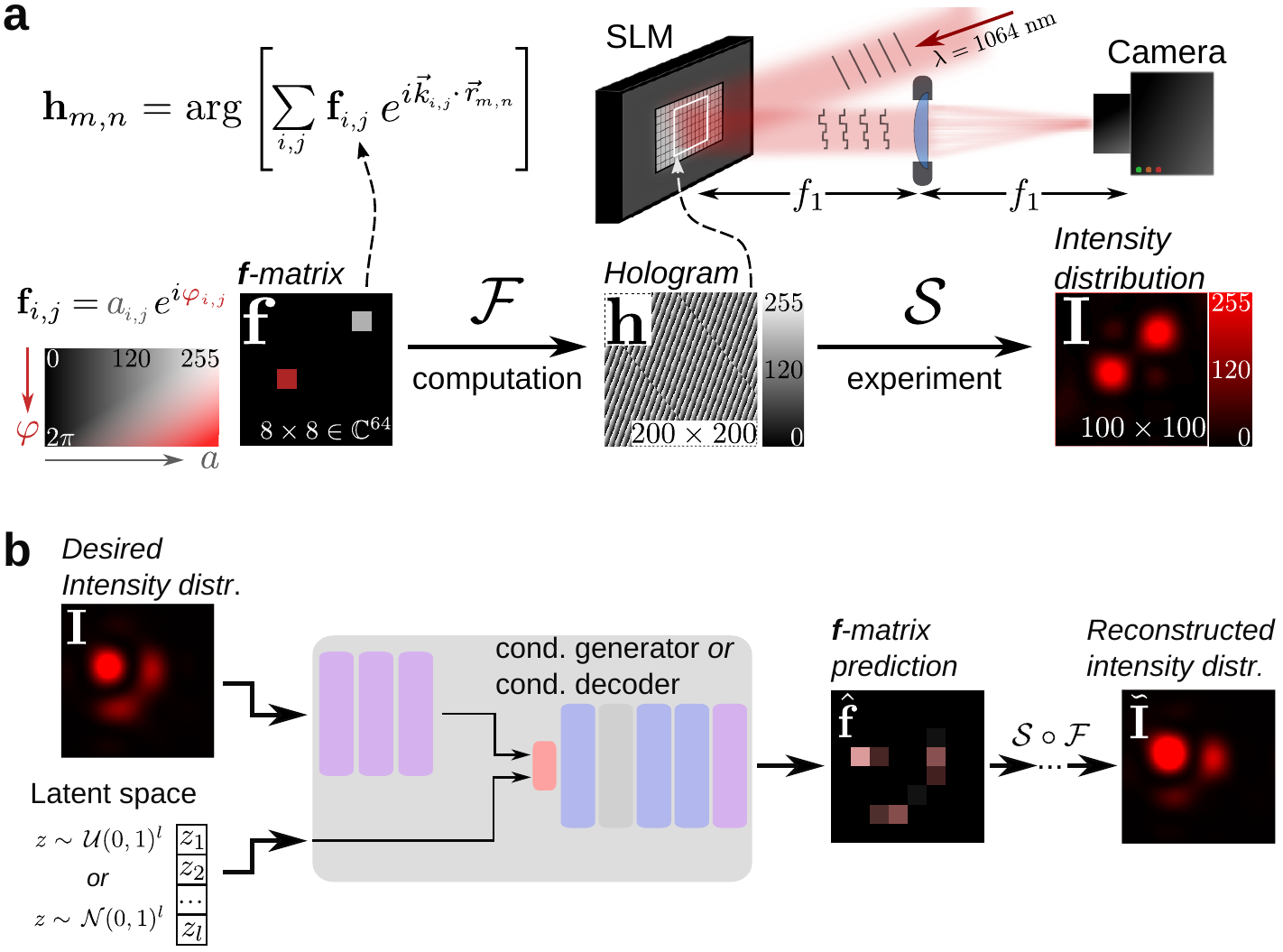}
    \caption{Flowchart of variables. We enhanced the contrast of pictures of laser intensity distributions for better visibility in both panels. {\bf a} Digital Holography: Initially, an $8\times 8$ matrix containing complex numbers is drawn or computed, which acts as a weighting matrix for a fixed set of 64 spatial frequencies $\vec{k}_{i,j}$. While the complex amplitude $a_{i,j}$ (greyscale) constitutes the weight, the complex argument $\varphi_{i,j}$ (redscale) plays the role of a phase-offset. 
    Using Eq.~\eqref{eq:blazed_transformation}, the $200\times 200$-shaped hologram $\mathbf{h}$ is computed from the $\mathbf{f}$-matrix and is displayed in the centre of the spatial light modulator (SLM) ($800\times600$ pixels). The SLM modulates the wavefront of the incoming laser beam in every of its active pixels (white frame). We then record the ensuing intensity distribution in the central $100\times 100$ pixels of the camera. In addition, the colour legend for each variable is given. {\bf b}~Phase-retrieval workflow: The desired intensity distribution is loaded into the generative model together with a latent space vector $z$. The model generates a $\mathbf{f}$-matrix estimate $\hat{\mathbf{f}}$ which is then used to create a hologram and then an intensity distribution~$\tilde{\mathbf{I}}$. For each target intensity, we redraw $z$-vectors several times and compare their performance in terms of their respectively achieved intensity error $E_\mathbf{I}=||\mathbf{I}-\tilde{\mathbf{I}}(z)||^2$. The coloured rectangles in the generator/decoder give a hint at the network architecture: The target intensity is subject to several convolutional layers (purple) before their output is concatenated with the latent space vector. The combined tensor then passes through a set of dense (blue) and dropout (grey) layers before the $\mathbf{f}$-matrix estimate is returned. }
  \label{fig:fig1}
\end{figure}

Historically, the idea of phase-retrieval algorithms goes back to Gerchberg and Saxton (GS)~\cite{Gerchberg1972}, who attempted to satisfy intensity constraints in two conjugate optical planes using repeated back-and-forth Fourier transformations.
The state of the field has been summarised by Di Leonardo \emph{et al.}~\cite{DiLeonardo2007}, who give a comprehensive review of existing phase-retrieval algorithms that can be used for SLM-based wave-shaping and develop a version of the GS algorithm to produce a grid of laser spots with an until then unprecedented degree of uniformity in spot intensity. 

More generally, challenges involved in phase-retrieval also appear in other contexts such as holographic imaging. Here, the phase problem manifests in a need for additional measurements or specialised numerical treatment~\cite{latychevskaia2007solution} in order to transform holograms to artefact-free images. Recently, Rivenson~\emph{et al.}~\cite{Rivenson2018} reported an image-reconstructing algorithm that aims to correct image artefacts which specifically arise due to missing phase information. The authors report using an appropriately trained neural-network-based model as a maximum-likelihood estimator for this purpose and present successful microscope image reconstructions of various biological tissues~\cite{Rivenson2018}.

In this study, we explore a data-driven, that is, assumption-free approach to holographic wave-shaping based on conditional deep generative models. Our approach is specifically tailored to finding approximate phase-only holograms for small two-dimensional intensity patterns with dimensions spanning just a few diffraction-limited-spot sizes. We compare several models with different objective functions, which are trained on the relationship between spatial hologram-frequencies, their respective phase-offset, and the resultant intensity patterns. Optical aberrations and misalignment-artefacts are thus automatically accounted for. Our model is able to (re)create laser patterns with complicated outlines beyond lines or points. Applications include optical tweezers-based studies of nano- and micron-scale thermodynamics, which often necessitate the creation of small intensity patterns with high precision~\cite{Grier2003, Gladrow2019}. However, the technique presented here is, in principle, applicable to larger patterns. The pattern size is limited only by available computational power or the size of the data-set used to train the neural networks that the model is composed~of.

In recent years, deep-neural-network-based machine learning has revolutionised the approach to a number of inference problems, such as image comprehension~\cite{LeCun1989, LeCun1998, Krizhevsky2012, Szegedy2014, He2015, Szegedy2016, Liu2018} or natural language processing~\cite{Collobert2011, Devlin2018, Radford2019}. During training, for instance on classification tasks, neural networks shape flows of increasingly abstract representations, from layer to layer, suppressing unnecessary features, while relevant ones are amplified~\cite{LeCun2015}. Under a supervised scheme, training refers to a minimisation of a loss function $\mathcal{L}$, e.g. $\mathcal{L}(\hat{y}, y)= || \hat{y}-y||_{l_2}^2$ over a labelled data set, such that the network output $\hat{y}(x)$ converges to the target output $y$ that corresponds to an input $x$~\cite{Bishop2006}. Crucially, naive attempts at fitting a network in this way to the data will only work for data relations that can be expressed by functions in the mathematical sense, i.e. by one-to-one- or many-to-one-type relations. Many real-world problems, such as the phase problem, however, contain one-to-many-type relations, are often highly oscillatory, and furthermore corrupted by noise at the input end. Any given output $y$ may thus correspond precisely or approximately to a range of inputs $\{x\}_y$. 

Here, we attempt to model such a relation using deep conditional generative models. Conditional generative models, such as conditional Generative Adversarial Networks (GAN) provide a means to approximate conditional data probability distributions~$\rho(x | y )$~\cite{Goodfellow2014, Mirza2014, Sohn2015} and thus offer a way to retain input variability for a given target output. Generative approaches to inverse problems have previously been used in molecular design~\cite{Sanchez-Lengeling2018} or computer tomography~\cite{Adler2018}. 

The idea is to invert the flow from holograms to intensity distributions depicted in Fig.~\ref{fig:fig1}{\bf a}, that is, model the distribution of holograms or precursors thereof conditioned on desired laser distributions as shown in panel {\bf b}. We compare model performances in terms of how well they (1) reconstruct previously measured, but unseen intensity distributions and (2) approximate synthetic distributions, for which a precisely reconstructing hologram might not exist. Specifically, we contrast performance of aforementioned conditional Generative-Adversarial Networks (cGAN) and Variational Autoencoder (cVAE) architectures with an expert system, which we introduce below. Our paper is structured as follows: We begin by defining the optics of digital holography, introduce our approach to generative modelling for inverse problems using an example, and then proceed to discuss the performance of our models for digital holography.

\section{Digital phase-only holography}
\label{sec:digital_holography}

Put simply, digital phase-only holograms are digital images where each pixel value encodes a jump (or delay) of the phase of incident wavefronts. In order to produce the desired pattern of intensity, the correct magnitude of phase change must be determined for each pixel. The spatial light modulator (SLM) sketched in Fig.~\ref{fig:fig1}{\bf a} is the physical apparatus which carries out this phase-shaping operation. The time-averaged incident laser beam in Fig.~\ref{fig:fig1}{\bf a} can be approximated as a plane wave, i.e. $E_0(\vec{r}) \propto e^{i\vec{v}\vec{r}}$ where $\vec{v}$ denotes the wave vector. The holograms displayed on the SLM affect the exponent, which is known as the phase of the wave $\varphi(\vec{r})$. The influence of the SLM in each of its pixels $(m,n)$ can be modelled by an unitary factor $p_{m,n}=e^{i \mathbf{h}_{m,n}}$ which is multiplied to the incident field $E_0$.  Downstream from the SLM, the laser beam passes a lens that is positioned in $2f$-configuration to the SLM and the camera. The lens Fourier-transforms the electrical field $E_\text{SLM}(\vec{r})$ into the field perceived by the camera $E_\text{Cam}(\vec{k})$~\cite{Mertz2009}. The intensity distribution recorded by the camera is then given by $I(\vec{k})= |E_{\text{Cam}}(\vec{k})|^2$, here discretised as $100\times 100$ image, $\mathbf{I}$.

Because of this Fourier-relationship, the hologram $\mathbf{h}$ that creates a single laser spot on the camera at $\vec{k}$ is given by a blazed grating, $\mathbf{h}_{m,n}= \textrm{arg}\left[  e^{i \vec{k}\cdot \vec{r}_{m,n}}\right]$. Generalising this relation to a grid with indices $i,j $ of multiple spots at fixed positions $\vec{k}_{i,j}$ then leads to the following equation,
\begin{align}
    \mathbf{h}_{m,n} &= \textrm{arg} \left[ \sum\limits_{i,j=1}^{N_p} \mathbf{f}_{i,j} e^{i \vec{k}_{i,j}\cdot \vec{r}_{m,n}}  \right] \label{eq:blazed_transformation}
\end{align}
where each element of the complex-valued matrix $\mathbf{f}_{i,j}=a_{i,j}e^{i\varphi_{i,j}}$ consists of a weighting factor (complex amplitude) $a_{i,j}$ and a phase-offset (complex argument) $\varphi_{i,j}$ for each spatial frequency $\vec{k}_{i,j}$ (see Fig.~\ref{fig:fig1}{\bf a}). In this paper, we restrict ourselves to $N_p = 8$ as a small proof of concept study for larger $\mathbf{f}$-matrices, which would lead to intensity patterns covering larger areas on the camera. Details on the optical setup can be found in the appendix Sec.~\ref{sec:exp_details}.

In Fig.~\ref{fig:fig2}{\bf a}, we give a few examples of the relationship between $\mathbf{f}$-matrices and ensuing intensity patterns. The first two examples, (i) and (ii), highlight the spatial correspondence of $\mathbf{f}$-matrix indices and laser spot positions. They also demonstrate the gain invariance of the argument function in Eq.~(\ref{eq:blazed_transformation}): any two $\mathbf{f}$-matrices $\mathbf{f}_1$, $\mathbf{f}_2$ which fulfil $\mathbf{f}_1 = c\mathbf{f}_2$ for a real $c>0$ will result in the same hologram $\mathbf{h}_1=\mathbf{h}_2$.    

The next two examples (iii), (iv) show a different invariance, which is of physical rather than mathematical nature: the phase invariance of light intensity which arises from the aforementioned relation between electrical fields and intensities, $I \propto | E | ^2$.
As we show in panel~{\bf b}, any two holograms $\mathbf{h}_1, \, \mathbf{h}_2 $, that differ only by a constant offset $\varphi_0$, i.e. $\mathbf{h}_1 = \left (\mathbf{h}_2 + \varphi_0\mathbf{1} \right)\, \textrm{mod} \, 2\pi$, will result in the approximately similar intensity distributions $\mathbf{I}_1\approx \mathbf{I}_2$. This can, for instance, be achieved by multiplying the $\mathbf{f}$-matrix by phase factor $\mathbf{f}_2= e^{i\varphi_0}\mathbf{f}_1$. However, the phase invariance in the examples (iii) and (iv) is a special case which can occur for $\mathbf{f}$-matrices with two non-zero elements (see Supp. Sec.~\ref{sec:phase_invariance_two}).

As the plot on the top in panel~{\bf b} shows, the phase invariance is not perfectly fulfilled in our experimental setup. The total light intensity of a single spot changes within $\pm 4\%$ as the four examples in panel~{\bf b} show. Moreover, phase invariances are rarely encountered for $\mathbf{f}$-matrices with more than one non-zero element as illustrated in the bottom two examples (v) and (vi) in panel~{\bf a}. In such a case, the intensity is almost only invariant in changes of global phase. In general, the interplay between different $\mathbf{f}$-matrix elements in terms of the resultant intensity is non-linear and non-local. In other words, a change in value of a certain $\mathbf{f}$-matrix element will impact the laser spots associated with all other elements, however distant in the matrix they are.

The task that we are concerned with in this paper is to find as many complex-valued matrices $\mathbf{f}$ as possible which will result in a given intensity distribution $\mathbf{I}$ or an approximation thereof. As shown in Fig.~\ref{fig:fig1}{\bf a}, we denote the two operations of computing the hologram $\mathbf{h}$ from $\mathbf{f}$, and obtaining the corresponding intensity $\mathbf{I}$ as $\mathcal{F}$ and $\mathcal{S}$. The main performance measure in this study is the intensity error $E_{\mathbf{I}, \tilde{\mathbf{I}}} = ||\mathbf{I} - \tilde{\mathbf{I}}||^2_{l_2} $ between instances of the target intensity $\mathbf{I}$ and the result of the setup operation $\tilde{\mathbf{I}}=\mathcal{S}\circ\mathcal{F}(\hat{\mathbf{f}})$ on a proposed $\mathbf{f}$-matrix. Due to the invariances described above, the function $\mathcal{S}\circ\mathcal{F}(\hat{\mathbf{f}})$ is not injective such that the associated inverse problem is ill-posed.

\begin{figure}
  \centering
    \includegraphics[width=154mm]{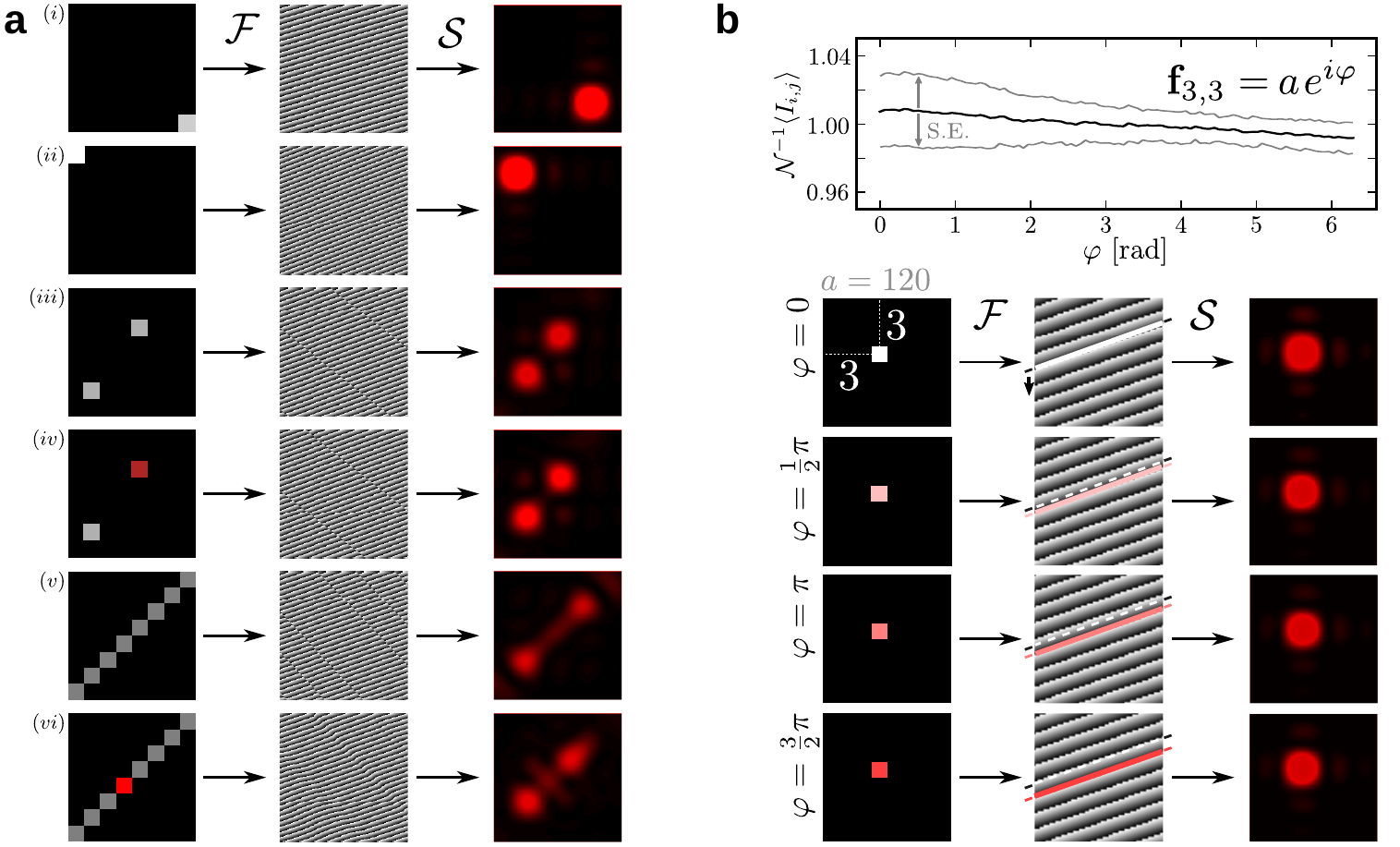}
  \label{fig:fig2}
  \caption{{\bf a)} Examples of $\mathbf{f}$-matrices, holograms $\mathbf{h}=\mathcal{F}(\mathbf{f})$, and resultant intensity distributions $\mathbf{I}=\mathcal{S}(\mathbf{h})$. Examples (i), (ii) highlight the spatial correspondence of $\mathbf{f}$-matrix indices and laser spots and the gain-invariance discussed in the main text. The examples in (iii), (iv) illustrate a special case of the phase-invariance, while the last two examples (v), (vi) show that this phase invariance breaks down for more than two non-zero $\mathbf{f}$-matrix elements.  {\bf b} The phase-invariance is a physical and only approximatively fulfilled as the plot on the top shows. The total intensity of a single spot varies within $4\%$ over all phase offsets. The pictures below exemplify the same phase invariance of single laser spots. All pictures of laser intensity distributions are contrast-enhanced for better visibility.}
\end{figure}

\section{Generative modelling and sampling strategy}
\label{sec:generative_modelling}
In this section, we introduce and motivate the concept of generative modelling, describe the sampling strategy used to construct the data sets, and describe how our models are trained. Futhermore, in order to illustrate our approach for non-experts, we solve a simple toy example of a non-invertible function in Sec.~\ref{sec:example_noisy_square}. The example also highlights the forward-interpolating properties of forward-trained conditional generative models, which we introduce in this paper.


\subsection{Model overview}
\label{sec:model_overview}
\paragraph{cGAN Model}
Ever since their discovery, generative models, such as GAN, have received attention due to their expressive abilities to create purely synthetic, but realistic pictures~\cite{Karras2018, Brock2018}. From a mathematical viewpoint, GAN are an approach to learning data probability distributions $\rho(x)$ in a generative way~\cite{Goodfellow2014, Nowozin2016}. 
More precisely, a GAN is a scheme aimed at training a neural network such that it is able to morph \emph{latent space} samples $z=\{z_1,\dots, z_l \}$ drawn from a standard probability distribution, such as the uniform distribution $\mathcal{U}^l(0,1)$, into samples $\tilde x=\{\tilde x_1, \dots, \tilde x_m \}$ that appear to be drawn from the data set $X$. This is achieved by training two networks in an antagonistic fashion: the first network, the discriminator, is trained to distinguish real samples from those that come from the generator. The generator, on the other hand, is trained to produce samples $\hat{x}$ from latent space vectors $z$ that satisfy the discriminator. 
Importantly, GAN can be trained in a conditional way, such that they learn to generate samples from an approximation to the conditional data distribution $\rho(x | y)$. Such a conditional data distribution could, for instance, originate from noisy function-input-output pairs $\{(x_1, y_1), \dots, (x_N, y_N) \}$ (see example in Sec.~\ref{sec:example_noisy_square}). Once trained, a conditional generator can generate multiple samples $\hat x(z, y)$ for a given $y$. In theory, this enables one to search for a latent space vector $z\in Z$, which minimises the forward-loss $E_y=\E{}_{y,z}\left[||y-f(\hat x(z, y)) ||^2_{l_2}\right]$ or at least retain invariant inputs $x$ which lead to the same~$y$. The discriminator, on the other hand, is obsolete after training.

Mathematically speaking, conditional GAN (cGAN) are trained over a data set distribution $\rho(\mathbf{f}, \mathbf{I})$ by performing stochastic gradient descent (SDG) on an approximate form of the following loss function~\cite{Goodfellow2014}
\begin{equation}
    E_\textrm{cGAN}(\{\mathbf{f}, \mathbf{I}\}; \phi, \theta) =  \E{}_{\mathbf{I}\sim \rho(\mathbf{I})}\left[\E{}_{\mathbf{f}\sim \rho(\cdot | \mathbf{I})}\left [ v\left( D_\phi (\mathbf{f}, \mathbf{I})\right) \right] + \E{}_{z\sim \mathcal{U}^l} \left [ w\left( D_\phi( G_\theta (z, \mathbf{I}),  \mathbf{I}) \right)\right] \right]+ \beta E_\textrm{rec}\left (\{\mathbf{f},\mathbf{I} \}; \theta \right)  \label{eq:gan_loss}
\end{equation} 
with respect to the parameters controlling the discriminator $D_\phi$ and generator $G_\theta$, $\phi$ and $\theta$. These parameter sets, $\theta$ and $\phi$, contain all trainable weights of the respective underlying neural network. The functions $v(\cdot)$ and $w(\cdot)$ in Eq.~(\ref{eq:gan_loss}) depend on the  type of GAN: in standard GAN, they are given by $v(x) = \textrm{log}(\sigma(x))$ and $w(x)=\textrm{log}(1-\sigma(x))$ with $\sigma(x)=(1+e^{-x})^{-1}$. In practice, the generator loss is often (here too) changed to $E_\textrm{gen} = \E{}_{z, \mathbf{I}}\left [-\textrm{log}\, D_\phi(G_\theta (z, \mathbf{I}),  \mathbf{I}) \right]$~\cite{Goodfellow2014}.
Expectations in Eq.~\eqref{eq:gan_loss} are estimated by running averages over small batches of data (mini-batching). The last term in Eq.~\eqref{eq:gan_loss} denotes a reconstruction loss, implemented throughout this paper as $E_\textrm{rec}(\{\mathbf{f}, \mathbf{I} \};  \theta) = \E{}_{\mathbf{I}, \mathbf{f}, z}\left[ ||\mathbf{f}- \hat{\mathbf{f}}(z, \mathbf{I})||^2_{l_2}\right]$. The hyperparameter $\beta$ controls its relative importance in relation to the rest of the GAN-loss. Since GAN are notoriously difficult to train, a number of normalisation-schemes have been proposed to prevent an otherwise likely loss of multimodality of the learned distribution $p_\theta(\hat{\mathbf{f}}|z, \mathbf{I})$, known as mode collapse~\cite{Salimans2016, Srivastava2017, Lee2019}. In this study, we normalised each layer in the discriminator $D_\phi$ using spectral normalisation~\cite{Miyato2018} (see Fig.~\ref{fig:discriminator_architecture}).

\paragraph{cVAE Model} In contrast to GAN, Variational Autoencoder (VAE) attempt to structure their latent space $Z$, which may simplify finding a suitable $z$-vector~\cite{Kingma2013}. More precisely, VAE enforce a particular distribution over $z\in Z$; typically a standard Gaussian is chosen. VAE, and their conditional form, cVAE, consist of two networks, a decoder and an encoder. The latter is trained to transform input variables $x$ and conditioned-on variables $y$ into a standard-normally distributed latent space vector $z\sim \mathcal{N}(0, \mathbf{1})$. The generative element in VAE, however, is the decoder. During training, the decoder generates samples $\hat x(z', y)$ from latent space vectors $z'$ drawn from the distribution specified by the encoder $\mathcal{N}(\mu_\textrm{enc}, \sigma_\textrm{enc}^2)$ and the conditioned-on variable $y$. Similar to the discriminator, the encoder becomes obsolete once training is completed. 

Mathematically speaking, the decoder controls the conditional probability $p_\theta:(z,\mathbf{I})\to \hat{\mathbf{f}}$, which maps from the latent space $z$ and desired intensity $\mathbf{I}$ into the $\mathbf{f}$-matrix space. The encoder, on the other hand, controls a latent space distribution $q_\phi(z | \mathbf{f})$ which is gradually trained towards a standard normal distribution. Once training is completed, the encoder can be removed; all latent space vectors are then drawn from $z \sim \mathcal{N}(0, \mathbf{1})$.
The training objective for both decoder and encoder can be combined into a single loss function given by
\begin{equation}
    E_\textrm{cVAE}(\{\mathbf{f}, \mathbf{I}\}; \phi, \theta) = \E{}_{\mathbf{I}\sim \rho(\mathbf{I}) }\left [\beta\E{}_{z \sim q_\phi (\cdot | \mathbf{f})} \left [\textrm{log} \, p_\theta(\mathbf{f}| z, \mathbf{I} ) \right]- \E{}_{\mathbf{f}\sim\rho(\cdot | \mathbf{I})}\left[D_\textrm{KL}\left(q_\phi(z | \mathbf{f}) || \mathcal{N}(0, \mathbf{1})\right)\right]\right],  \label{eq:vae_loss} 
\end{equation}
where $D_\textrm{KL}(a || b)$ represents the Kullback-Leibler divergence between the two arguments. The idea behind Eq.~\eqref{eq:vae_loss} is to optimise a lower bound for the probability of the distribution controlled by the decoder $p_\theta(\mathbf{f} | z, \mathbf{I})$ of matching the actual, but unknown distribution $\rho(\mathbf{f} | \mathbf{I})$~\cite{Kingma2013}. In this study, we implement the first term as the reconstruction loss in $\mathbf{f}$-matrix space with $\beta$ again controlling its relative importance. Throughout this study, we set $\beta=1$. 
The second term in Eq.~\eqref{eq:vae_loss} enforces a particular distribution of the latent space variable $z\in Z$ during the training process.

\paragraph{cVAE + forw. loss} Models that are trained only by minimising the $\mathbf{f}$-error will not be able to correctly interpolate in intensity space. We attempt to remedy this by adding a forward network $U_\xi$ during training, which constructs a differentiable representation of the forward operation $\mathcal{F}\circ\mathcal{S}: \mathbf{f} \to \mathbf{I}$.

The forward network $U_\xi:\mathbf{f} \to \hat{\mathbf{I}}$ with parameters $\xi$ is pretrained on the data set using a forward error, $E_{\mathbf{I}, \hat{\mathbf{I}} }  = ||\hat{\mathbf{I}} - \mathbf{I} ||_{l_2}^2$. We note the difference in notation of $\hat{I}$ in contrast to the actual setup output $\tilde{\mathbf{I}}=\mathcal{S}\circ\mathcal{F}(\hat{\mathbf{f}})$.

Once pretraining is completed, the forward network is used to evaluate the $\mathbf{f}$-matrix candidates $\hat{\mathbf{f}}$ proposed by the generative models in terms of their likely intensity pattern $\hat{\hat{\mathbf{I}}}=U_\xi(\hat{\mathbf{f}})$ and pass on $l_2$-loss from the intensity plane. During the actual training, we continue to minimise $E_{\mathbf{I}, \hat{\mathbf{I}}}$ with respect to $\xi$ in a separate training step as discussed in the example. We emphasise that the forward-gradient descent steps of the VAE are taken only with respect to the en- and decoder parameters $\phi$, $\theta$, not~$\xi$.

In principle, one can augment the loss function of any conditional generative model in this way. We here choose the cVAE. The resulting forward-cVAE is updated using the following loss function 
\begin{equation}
    E_\textrm{VAE + forw.loss }(\{\mathbf{f}, \mathbf{I}\}; \xi, \phi, \theta) =  \E{}_{\mathbf{I}\sim\rho(\mathbf{I})}\left[\alpha\E{}_{\mathbf{f}\sim \rho(\cdot | \mathbf{I})} ||U_\xi(\hat{\mathbf{f}}) - \mathbf{I} ||^2_{l_2}\right] + E_\textrm{VAE}(\{\mathbf{f}, \mathbf{I}\}; \phi, \theta). \label{eq:vae_forward_loss} 
\end{equation}
where the new hyperparameter $\alpha$ controls the relative strength of the forward loss. We explain this type of model in further detail in the example in Sec.~\ref{sec:example_noisy_square}. We found it beneficial to train the forward network $U_\xi$ under spectral normalisation, similar to the discriminator in our cGAN model. A thorough investigation of the benefit of this, however, is beyond the scope of this study.

\paragraph{Expert system} Our fourth model is not based on neural networks and serves as a baseline against which we can compare the performance of our models. It is an attempt to find a direct relation between the value of $\mathbf{f}$-matrix elements and laser spot strengths in the intensity distribution~$\mathbf{I}$.

Using a peak-finding technique, we inferred affine transformations, $i = a_y I^\textrm{peak}_y + b_y$ and $j = a_x I^\textrm{peak}_x + b_x$, which transform peak-positions of isolated laser spots in the intensity plane into $\mathbf{f}$-matrix indices $i,j$. The absolute values of the elements at these indices can now be set to some value, e.g. $|\hat{\mathbf{f}}_{i_,j}| = c\cdot I^\textrm{peak}_{y,x}$, where $c$ is a linear scaling factor, which we set to $c=1$. The peak-finder requires a threshold to binarise the image, which must be seen as a model parameter. In all experiments, we used thr=$0.5\cdot \textrm{max}(\mathbf{I})$.
The peak position is then inferred using a centroid from the non-binarised image within a region of interest determined in the binarised image. The expert system therefore produces estimates for each $\mathbf{f}$-matrix element based on the value of the respective laser spot without reference to other spots.

Importantly, for the phase $\varphi_{i,j}$ of the $\hat{\mathbf{f}}$-matrix elements, we do not have a model. We therefore resort to drawing random phases $\varphi_{i,j}\in [0, 2\pi)$ for the non-zero elements in $\hat{\mathbf{f}}$.

\subsection{Example: Inverting the noisy square function}
\label{sec:example_noisy_square}
\begin{figure}
    \centering
    \includegraphics[width=154mm]{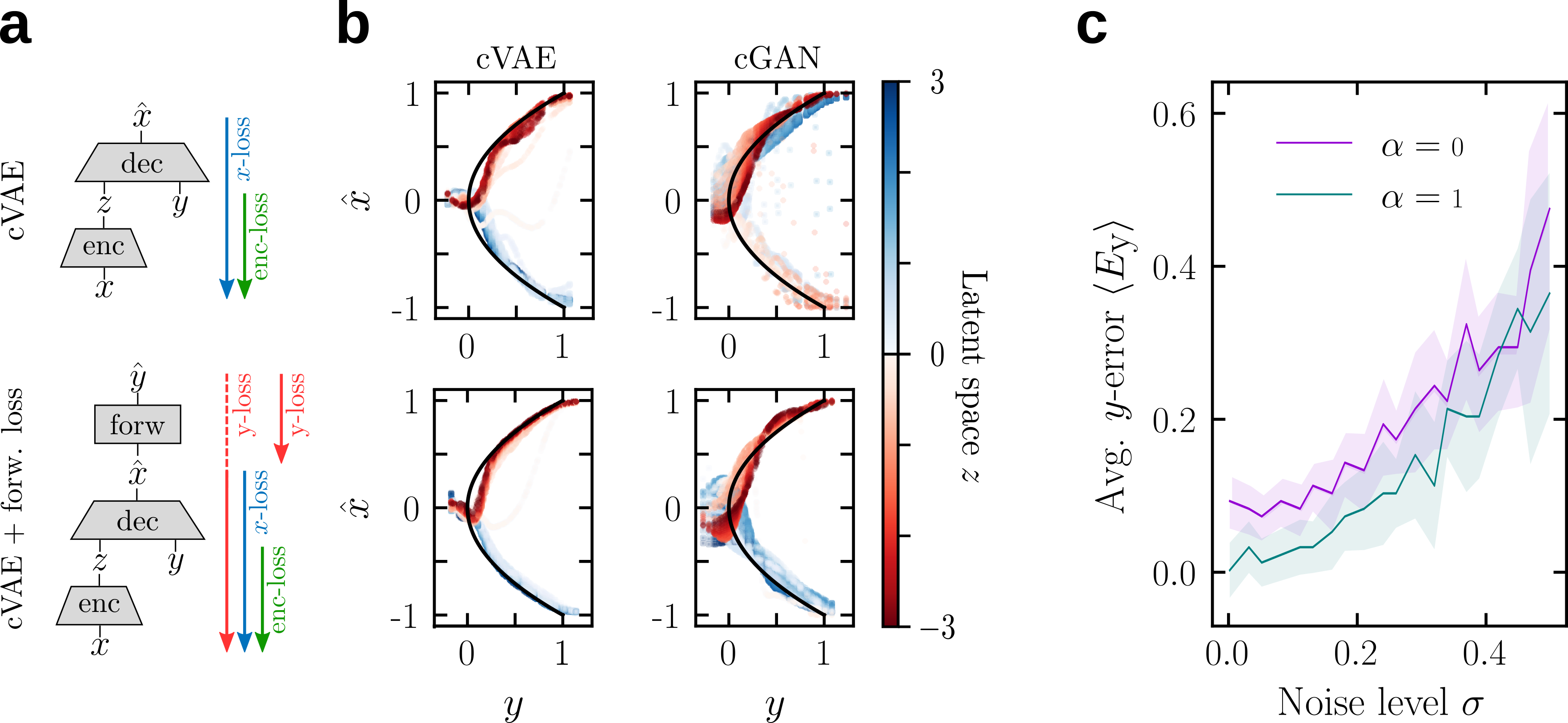}
    \caption{Inverting a noisy square function $f(x)=x^2 + \xi$ using conditional generative models. {\bf a} Sketch of the gradient flow during training for standard cVAE (top) and cVAE with forward loss (bottom).  {\bf b} Examples of cVAE (left column) and cGAN (right column) trained without (top row) and with (bottom row) forward loss for a noise level of $\sigma = 0.05$. {\bf b} Average $y$-reconstruction error of a cVAE trained without (violet) and with (teal) forward loss as a function of the level of input noise. }
    \label{fig:inverse_example_figure}
\end{figure}

We provide a simple example in order to illustrate the concept of generative modelling and the idea of using a forward-loss, which we introduced in the third model in the previous section. The task here is to approximate the inverse, $f^{-1}$, of the noisy square function $f(x) = y =  x^2 + \xi$ with $\xi \sim \mathcal(0, \sigma^2)$ from a data set of pairs $\{(y_1,x_1), (y_2, x_2), ..., (y_N, x_N)\}$ for $x \in [0,1]$. In this example, the inputs $x$ will take up the role of the $\mathbf{f}$-matrix in the actual study, while the output $y$ assumes the role of the intensity matrix $\mathbf{I}$. In Fig.~\ref{fig:inverse_example_figure}{\bf a}, we sketch the architecture of a cVAE during training. The various loss functions are $l_2$-difference norms in the $x$ (blue) or $y$ plane (red) with the exception of the encoder loss (green), which is given by the Kullback-Leibler divergence in Eq.~\eqref{eq:vae_loss}. In contrast to the classical implementation of cVAEs, we found it sufficient to provide the encoder with the input $x$, instead of the tuple $(x, y)$ as indicated in panel~{\bf a}.

In Fig.~\ref{fig:inverse_example_figure}{\bf b}, we show inverse interpolations using a cVAE (left) and a cGAN model (right) with a single latent space variable $z$, which is encoded by colour (see colour bar). Once trained, the models are presented with a target output value $y$ and a randomly chosen latent space value $z$ and are asked to predict an input $\hat{x}$ that leads to $y$. While the two branches associated with negative and positive latent space values in cVAEs are visibly disentangled, they are interfusing in the case of cGAN~\cite{higgins2017beta}. This is an example of the effect of the encoder loss in Eq.~\eqref{eq:vae_loss}.

In the lower row, we give an example of a forward loss: to this end, we add another network $U_\xi$ which is trained on the noisy forward relationship $x \to y$ (third model in Sec.~\ref{sec:model_overview}). Importantly, as we mention before, the forward network is trained independently from the $x$-estimator (separate short red arrow in the bottom row of Fig.~\ref{fig:inverse_example_figure}{\bf a}) and only used to pass on $l_2$-error from the $y$ plane (red-dashed line). The generative model is then trained by descending the forward loss $E_\textrm{forw}(\{ y, z\} )=\E{}_{y, x}||y- U_\xi(\hat{ x}(z, y))||^2_{l_2}$ in addition to the $x$-loss. As before, the relative strength of the forward loss is measured by $\alpha$. In panel {\bf c}, we show that such a network architecture can interpolate in $y$-space ($\alpha = 1$), which in our case leads to a statistical reduction in the $y$-error, $E_y = || y - f(\hat{x}) ||_{l_2}^2$. This forward interpolating property is relevant for our holography-predicting generative models, especially when presented with synthetic intensity distributions.

\subsection{Sampling strategy}
\label{sec:sampling_approach}
\begin{figure}
  \centering
    \includegraphics[width=154mm]{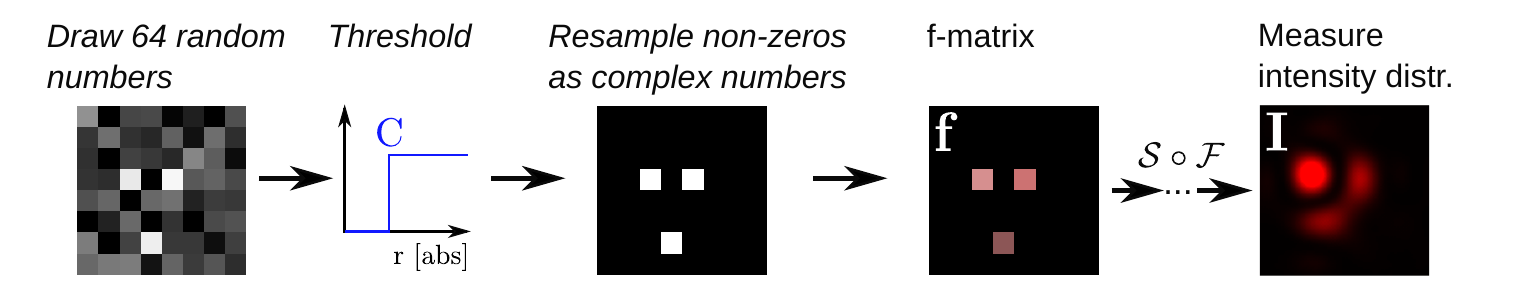}
    \caption{Data set sampling workflow used to create training and test data sets. The workflow is designed to result in sparse $\mathbf{f}$-matrices, which result in simpler, more concentrated intensity distributions.}
  \label{fig:fig3}
\end{figure} 

A deep-learning approach to digital holography necessitates compiling large training data sets of $\mathbf{f}$-matrices and corresponding intensity distributions~$\mathbf{I}$. Since we do not have fixed sets of training, validation, and test data at our disposal, the sampling strategy we use to compile these data sets is relevant. It is crucial that the data set must be constructed in such a way that it likely encompasses relevant intensity patterns. For instance, simply drawing 64 complex numbers at random would lead to non-sparse $\mathbf{f}$-matrices, which typically result in low-light intensity distributions, due to hologram overloading. We therefore adopt a different approach sketched in Fig.~\ref{fig:fig3} in which we draw 64 uniformly distributed random numbers and compare them with a predefined sparseness threshold $C$. We then redraw the radius and phase of those complex numbers $\mathbf{f}_{i,j}$ where the threshold was surpassed. The radii and phases of all other $\mathbf{f}$-matrix elements are set to zero. This is a simple way of achieving a Bernoulli-distributed preselection of spatial frequencies. 
In order to explicitly explore phase-intensity relation, we redraw the phases of each sample three times and measure the resultant intensity distribution. The data set compiled in this way consists of 150,000 ($\mathbf{f}$, $\mathbf{I})$-pairs (50,000 different amplitude patterns).  

\section{Results}
\label{sec:results}
In the following, we discuss the performance of our models on a test data set and the creation of as well as the model performance on a purely synthetic intensity data set. In both cases, the workflow follows the sketch in Fig.~\ref{fig:fig1}{\bf b}, with the addition that for each intensity distribution $\mathbf{I}$, we draw multiple latent space vectors $z$ and measure the ensuing intensity. Throughout this study, we compare the measured intensities $\tilde{\mathbf{I}}_1, \dots, \tilde{\mathbf{I}}_5$ for five different latent space vectors $z_1, \dots, z_5$ and report the lowest intensity error. Similarly for the expert system, we redraw the phase offsets of non-zero $\hat{\mathbf{f}}$-matrix elements five times.

The architecture of decoder and generator networks indicated in Fig.~\ref{fig:fig1}{\bf b} is adhered to by all our generative models: The translation-invariant properties of the combined mapping $\mathcal{F}\circ\mathcal{S}:\mathbf{f}\to \mathbf{I}$ (see Fig.~\ref{fig:fig2}{\bf a}) invite the use of convolutional layers. We therefore chose to send the target intensity $\mathbf{I}$ through several convolutional layers which reduce its initial $100\times100$ size down to a $8\times8\times 8$ tensor, before concatenating the reduced tensor with the latent space vector $z$. The combined tensor then proceeds through a set of dense layers, interleaved by a dropout layer~\cite{srivastava2014dropout} in order to reduce the risk of overfitting. All layers in the generator/decoder networks are subject to batch normalisation~\cite{ioffe2015batch}. Details of network architectures and training procedure are summarised in the appendix Sec.~\ref{sec:architecture}.

\subsection{Reconstructing previously measured intensities (test data set)}
\label{sec:reconstructing_intensities}
We compare the generalising abilities of all four models described in Sec.~\ref{sec:model_overview} on a test data set containing 500 elements, sampled according to the recipe described in Sec.~\ref{sec:sampling_approach}. In this data set, the model is simply asked to reconstruct intensity distributions that have been recorded before. Importantly, this guarantees the existence of a hologram disregarding measurement error and mechanical drift.

\begin{figure}
    \centering
    \includegraphics[width=154mm]{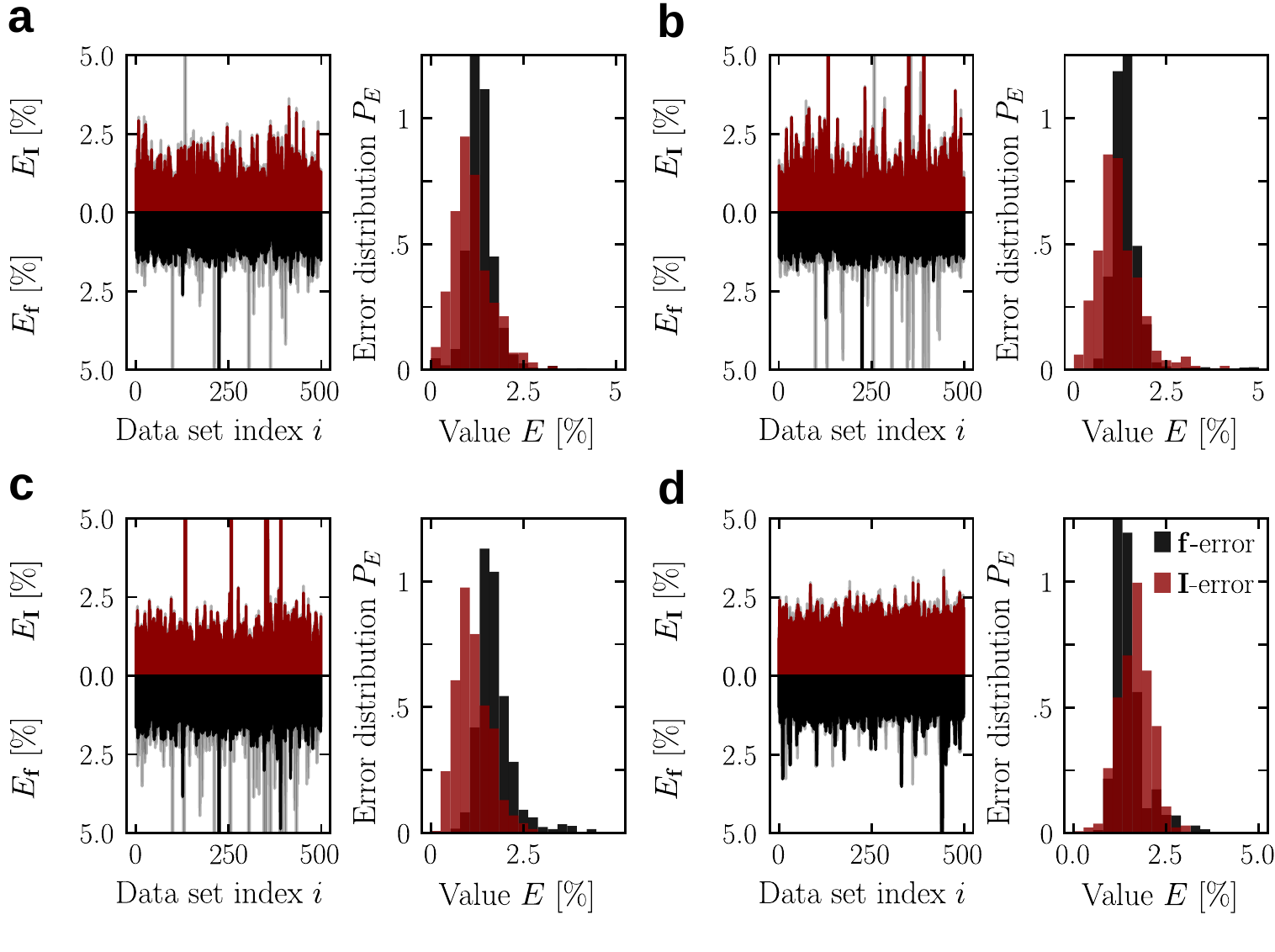}
    \caption{Overview of test-set intensity- and $\mathbf{f}$-error ($E_\mathbf{I}$, $E_\mathbf{f}$) achieved by our models. Both errors are normalised by the total target intensity $\sum_{i,j} \mathbf{I}_{i,j}$ and the absolute sum of the original $\mathbf{f}$-matrix, $\sum_{i,j} |\mathbf{f}_{i,j}|$ respectively. In the left panels, the coloured bars indicate the minimal error, while the grey bars show the maximum error measured over all five latent space redraws. {\bf a} cVAE, {\bf b} cGAN, {\bf c} cVAE with forward loss, and {\bf d} the expert system. }
    \label{fig:validation_overview}
\end{figure}

The average intensity errors $\langle E_y \rangle$ are summarised in the second column in Tbl.~\ref{tab:int_error_overview}. A detailed overview and histogram of the intensity and $\mathbf{f}$-matrix error is given in Fig.~\ref{fig:validation_overview}. In order to give an impression of the variance in the intensity- and the $\mathbf{f}$-matrix plane achieved by redrawing latent space vectors, we plot the highest $\mathbf{f}$-error in grey in the left panels of Fig.~\ref{fig:validation_overview}. 

\begin{table}[]
    \centering
    \begin{tabular}{l c c}
    Model & Test Set $\langle E_\mathbf{I} \rangle$ [\%] & Synthetic Set $\langle E_\mathbf{I} \rangle$ [\%] \\
    \hline
    \hline
        cVAE  & $1.16 \pm 0.04$   &   $3.43 \pm 0.07 $\\
        cGAN & $1.27 \pm 0.05$ & $3.48 \pm 0.07$ \\
        cVAE + forw. loss & $1.36 \pm 0.14$ & $2.56 \pm 0.06$\\
        Expert system & $1.65 \pm 0.08$& $3.4 \pm 0.1$ \\
        \hline
    \end{tabular}
    \caption{Overview of mean-intensity errors achieved by our models on the test as well as the synthetic data set.}
    \label{tab:int_error_overview}
\end{table}
As both the table and the figure show, the standard cVAE achieves the lowest relative intensity error, followed by the cGAN model, the cVAE trained with intensity loss, and the expert system. All three generative models exhibit $\mathbf{f}$-matrix variability for different latent space vectors as the spread of the $\mathbf{f}$-error $E_\mathbf{f}$ in Fig.~\ref{fig:validation_overview} shows. The spread in intensity-error $E_\mathbf{I}$ is significantly lower, in accord with expectations: the models are trained to produce $\mathbf{f}$-matrix candidates for a given intensity $\mathbf{I}$. This observation appears to be particularly the case for the cVAE with forward loss in panel {\bf c}. The models do not, however, appear to have retained the full phase spectrum for single $\mathbf{f}$-matrix elements. A possible cause might be the only approximative nature of the phase-invariance shown in Fig.~\ref{fig:fig2}{\bf b}. It is equally possible, that the models have not seen the necessary amount of examples of this invariance in the data set despite the explicit phase-resampling step discussed in the previous section.

In any case, relying solely on measured intensity errors $E_\mathbf{I}$ can be deceiving, since the precise laser position at the spatial frequencies $\vec{k}_{i,j}$ measured during acquisition of the training and test data set may have shifted slightly over the course of time. Indeed, the precise position of laser patterns on the camera exhibits a day-to-day variability of around one to three camera pixels, presumably due to thermal variability affecting optical components. In order to keep this variability at a minimum we 'preheated' the setup by running the laser for 30 min prior to recording the error. Furthermore, we adjusted the camera region of interest by 1-3 pixels to counter any shift in spot position.

However, despite these measures, we record a floor of $0.4\%$ of intensity error, when we reuse the exact same holograms from the test set. All relevant experimental details of the optical setup are summarised in the appendix~\ref{sec:exp_details}.

\begin{figure}
    \centering
    \includegraphics[width=154mm]{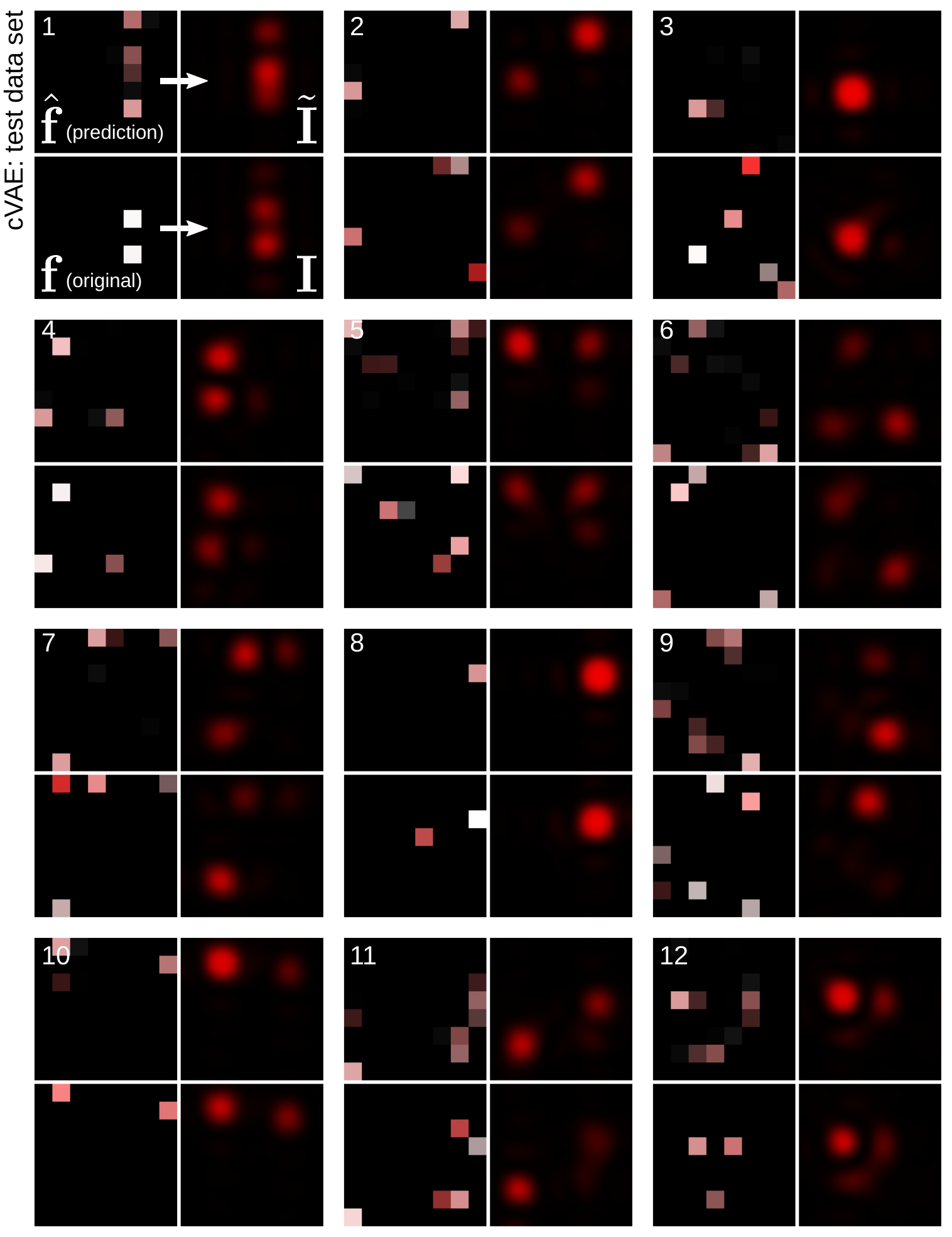}
    \caption{cVAE test data set examples. The colour code for the $\mathbf{f}$-matrices and the intensities is given in Fig.~\ref{fig:fig1}. For each example, we present the intensity and $\mathbf{f}$-matrix prediction that achieved the lowest intensity-error over all latent space draws.}
    \label{fig:cVAE_valid_examples}
\end{figure}

\begin{figure}
    \centering
    \includegraphics[width=154mm]{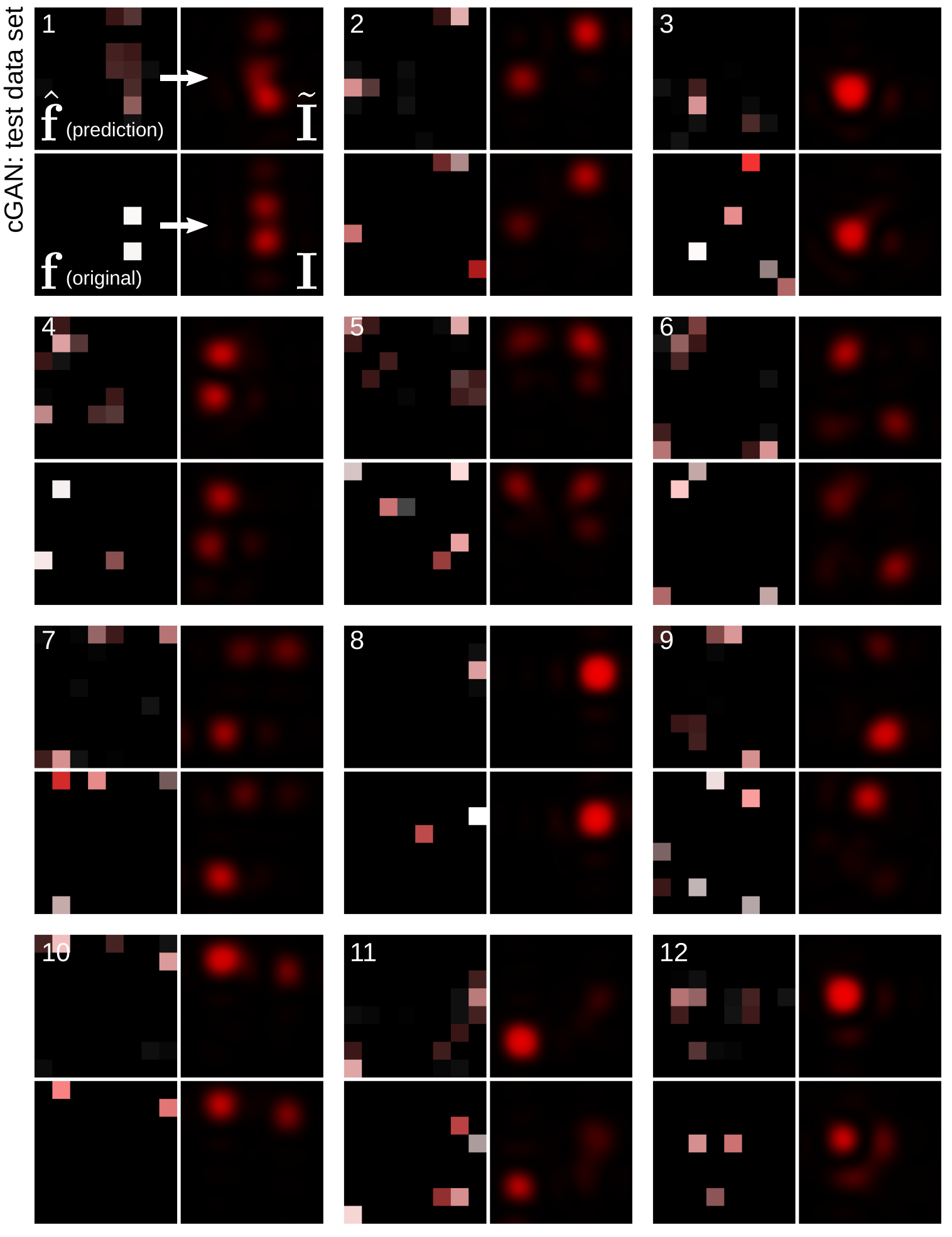}
    \caption{cGAN test data set examples.}
    \label{fig:cGAN_valid_examples}
\end{figure}

\begin{figure}
    \centering
    \includegraphics[width=154mm]{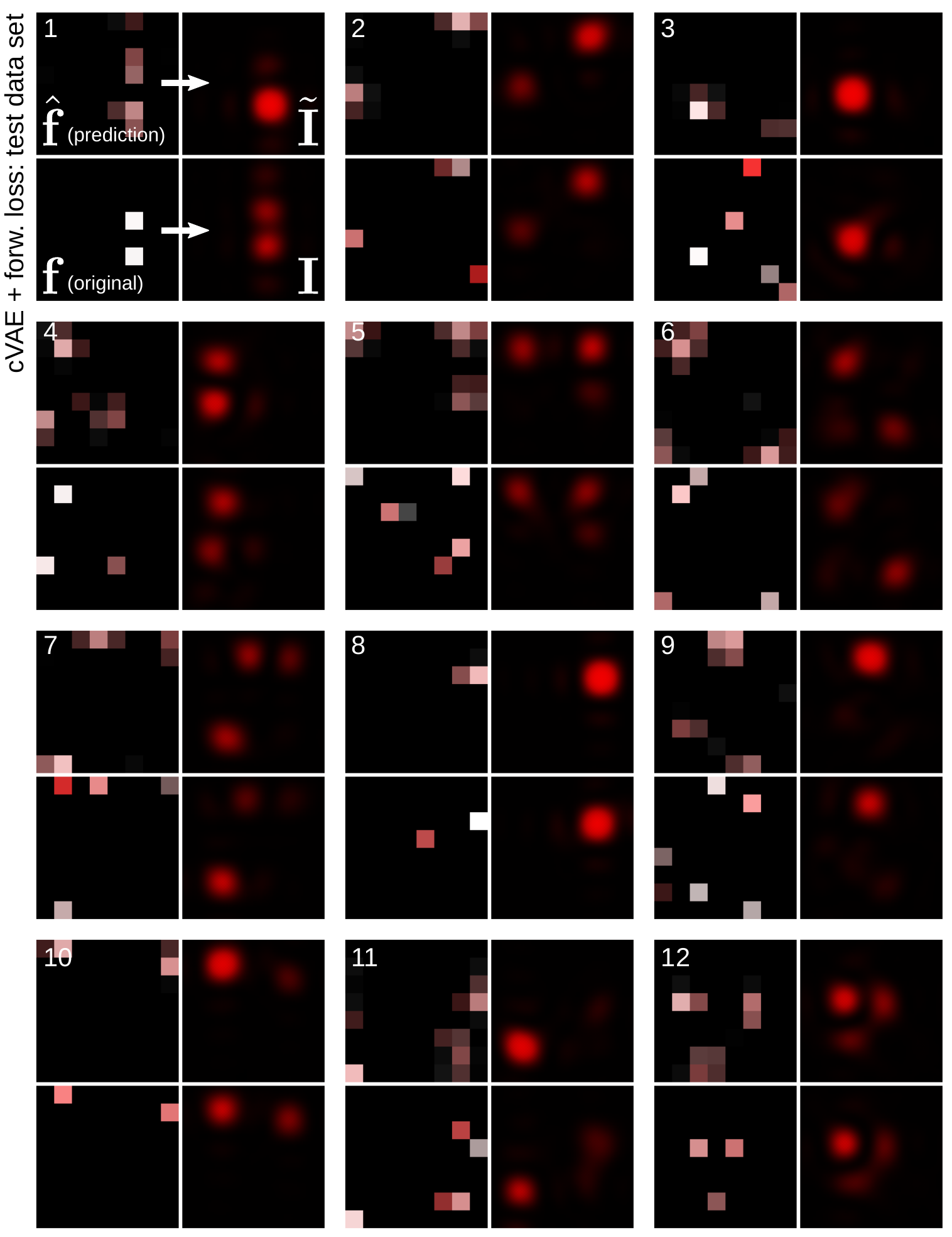}
    \caption{cVAE with forward loss test data set examples.}
    \label{fig:cVAE_forward_valid_examples}
\end{figure}

\begin{figure}
    \centering
    \includegraphics[width=154mm]{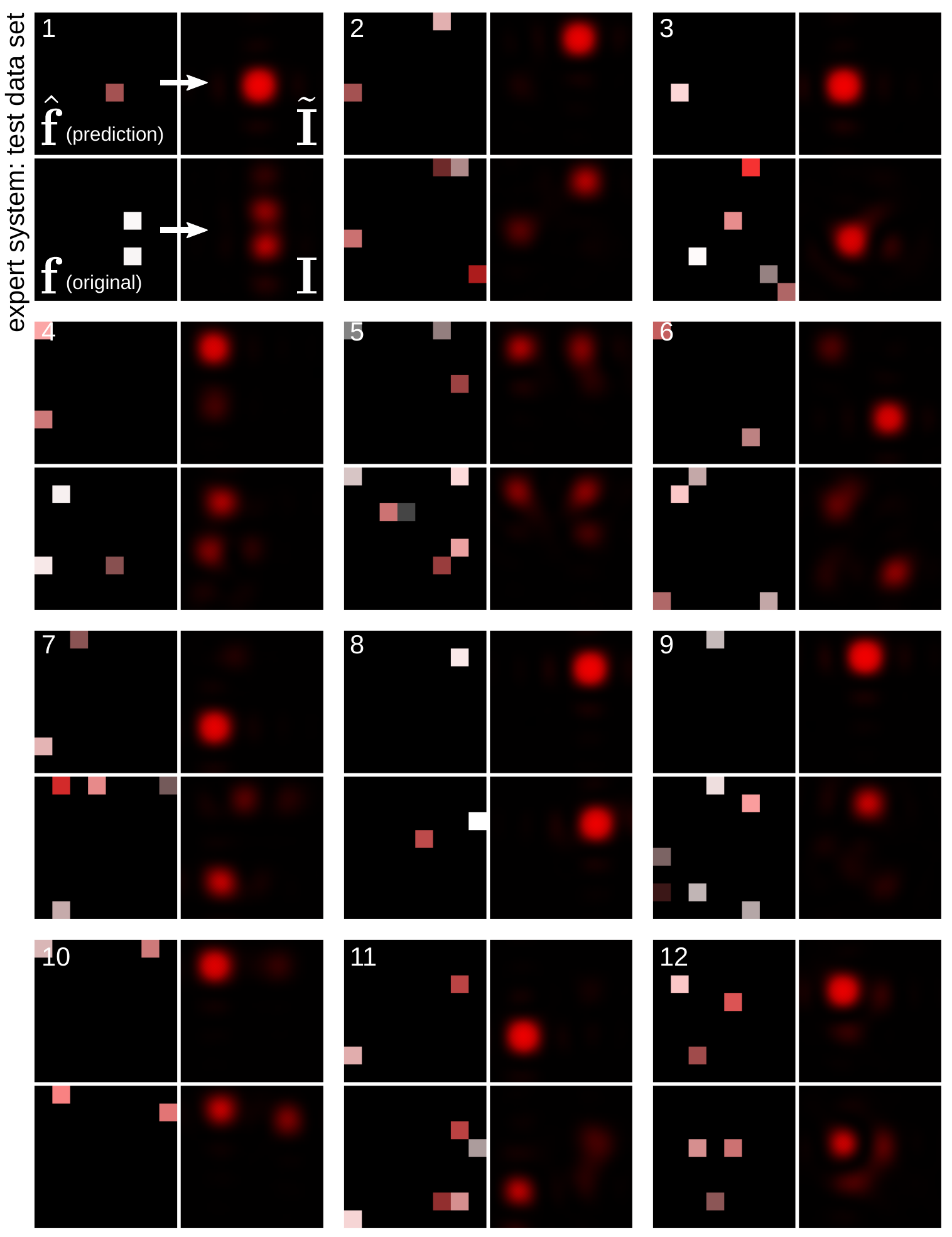}
    \caption{Expert system test data set examples.}
    \label{fig:expert_valid_examples}
\end{figure}

A visual inspection of shapes and positions of reconstructed intensities is therefore still a valuable measure of model performance. We give an extensive collection of 12 test-set examples for each model in Figs.~\ref{fig:cVAE_valid_examples}-\ref{fig:expert_valid_examples}. In order to avoid any selection bias, we chose the last 12 examples in the test set. In these figures, the intensity distributions are not contrast-enhanced, since otherwise the impression would be distorted. The cVAE model (Fig.~\ref{fig:cVAE_valid_examples}) exhibits an excellent ability to reconstruct even more complicated shapes such as  example 12, in which a quarter-circle shape surrounds a central spot. The cVAE model that was trained with a forward loss manages to reproduce this structure too.
By contrast, in example 9, the standard cVAE weighs the spot on the top too strongly. The cVAE with forward loss, on the other hand, appears to underestimate the weight of the top spot in example 1.

The results of the expert system are shown in Fig.~\ref{fig:expert_valid_examples}. Its performance appears inconsistent as it manages to reproduce both spots in example 6, but fails to do so in example 1, 2, 7, and 11. 

\subsection{Approximating synthetic intensity distributions}
\label{sec:approximating_synthetic_intensity_distributions}
The goal of this study is to construct models that are capable of approximating synthetic target intensities that have not been previously measured. Indeed, in many applications of digital holography, target intensity patterns will vary across a wide range of possible patterns and likely include 'impossible patterns'. From an applied perspective, it is often not \emph{a priori} clear whether a hologram that results precisely in the target intensity even exists. A desirable property of models for digital holography is therefore to propose $\mathbf{f}$-matrices or holograms $\mathbf{h}$, that result in an intensity $\tilde{\mathbf{I}}$ that is as close as possible to the target~$\mathbf{I}$. However, the models discussed here do not interpolate in $\mathbf{I}$-space, but $\mathbf{f}$-matrix space with the exception of the third model, the cVAE trained with an intensity loss.

In the following, we test the three generative models and the expert system on 500 entirely synthetic intensity distributions $\mathbf{I}$. We create these distributions using a mixture of Gaussians with a variable number of spots $0< k\leq N_p$, with peak-positions $(\mu^{(k)}_x, \mu^{(k)}_y)$, amplitudes $A_k$, variances $(\textrm{var}^{(k)}_x, \textrm{var}^{(k)}_y)$, and $x$-$y$-covariances $\textrm{cov}^{(k)}_{x,y}$,
\begin{equation}
    \mathbf{I}_\textrm{synth}(x,y) = \sum\limits_{k=0}^{N_p} A^{(k)} e^{-\frac{1}{2} \frac{\textrm{var}^{(k)}_y(x-\mu^{(k)}_x)^2-2\textrm{cov}^{(k)}_{x,y}(x-\mu^{(k)}_x)(y-\mu^{(k)}_y)+\textrm{var}^{(k)}_x(y-\mu^{(k)}_y)^2 }{\textrm{var}^{(k)}_x\textrm{var}^{(k)}_y-\textrm{cov}^{(k)}_{x,y}\textrm{cov}^{(k)}_{x,y}}}. \label{eq:mixt_gaussian}
\end{equation}
All parameters are uniformly distributed. However, in an attempt to create meaningful intensity distributions, we require peak positions $(\mu_x, \mu_y)$ in Eq.~\eqref{eq:mixt_gaussian} to respect a margin to the image boundaries of 20 pixels and place bounds on the variances, $50 \leq \textrm{var} \leq 65$. Furthermore, the maximum achievable intensity for each spot amplitude $A^{(k)}$ is bound from above by $A_\textrm{max} = 300/\sqrt{N_p}$, where $N_p$ is a discrete uniformly-distributed random number drawn initially $N_p \in \{1, \dots, 5\}$. The spot intensity $A^{(k)}$ is also bounded from below by $A_\textrm{min} =40$. Despite these restrictions, the intensity distributions might still be impossible to precisely realise: for instance, spot amplitudes and spot-shape variances are not independent parameters. The intensity distributions created here are thus outside of the ensemble of the training data set and put the ability of the conditional generative models to generalise to a test.

\begin{figure}
    \centering
    \includegraphics[width=154mm]{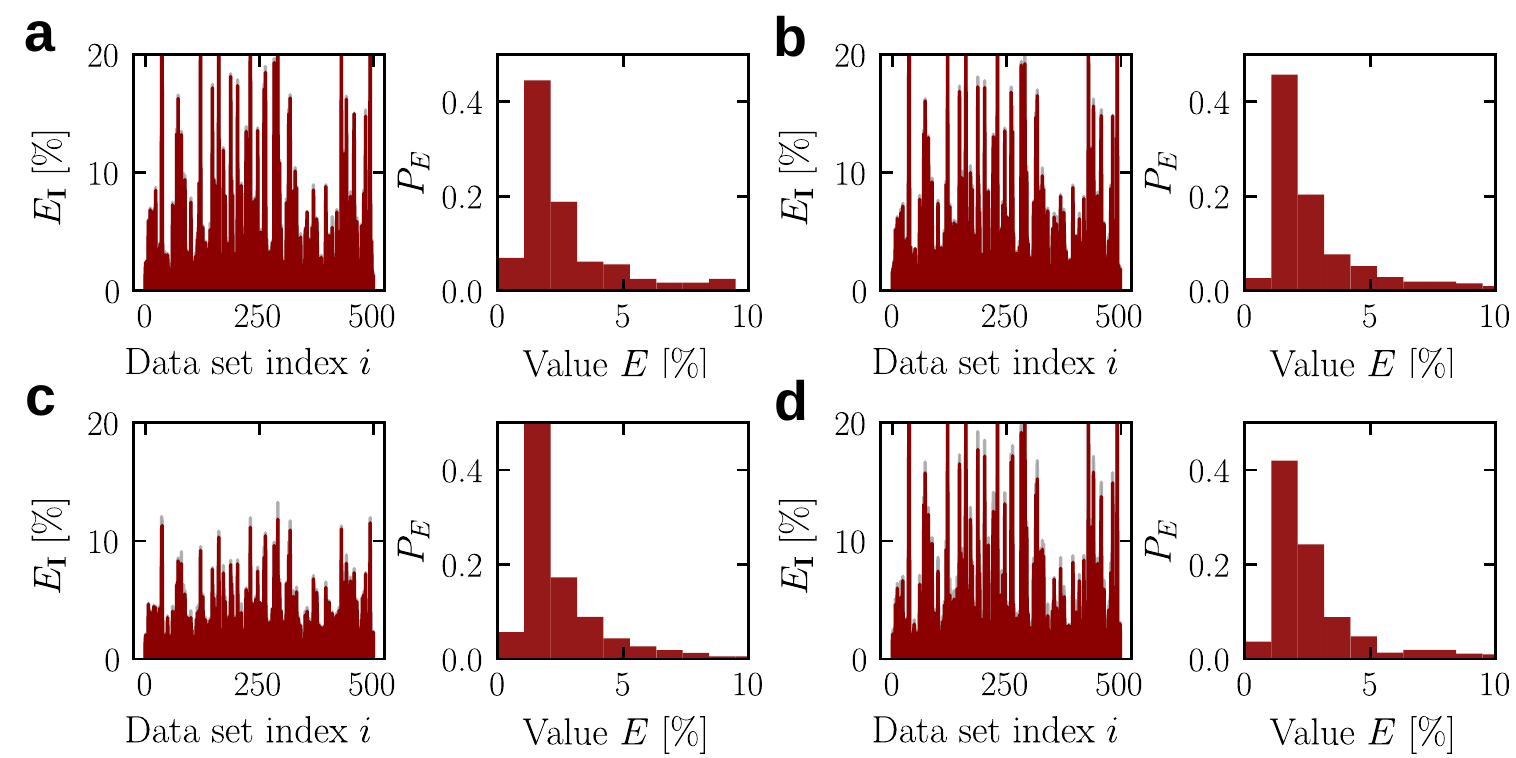}
    \caption{Overview of the intensity error achieved by our models on the synthetic data set. Similar to the panels on the left in Fig.~\ref{fig:validation_overview}, the latent space draw that resulted in the minimum intensity error is shown in red, while the draw resulting in the maximum error is shown in grey. {\bf a} cVAE, {\bf b} cGAN, {\bf c} cVAE with forward loss, and {\bf d} expert system. }
    \label{fig:testset_overview}
\end{figure}

\begin{figure}
    \centering
    \includegraphics[width=154mm]{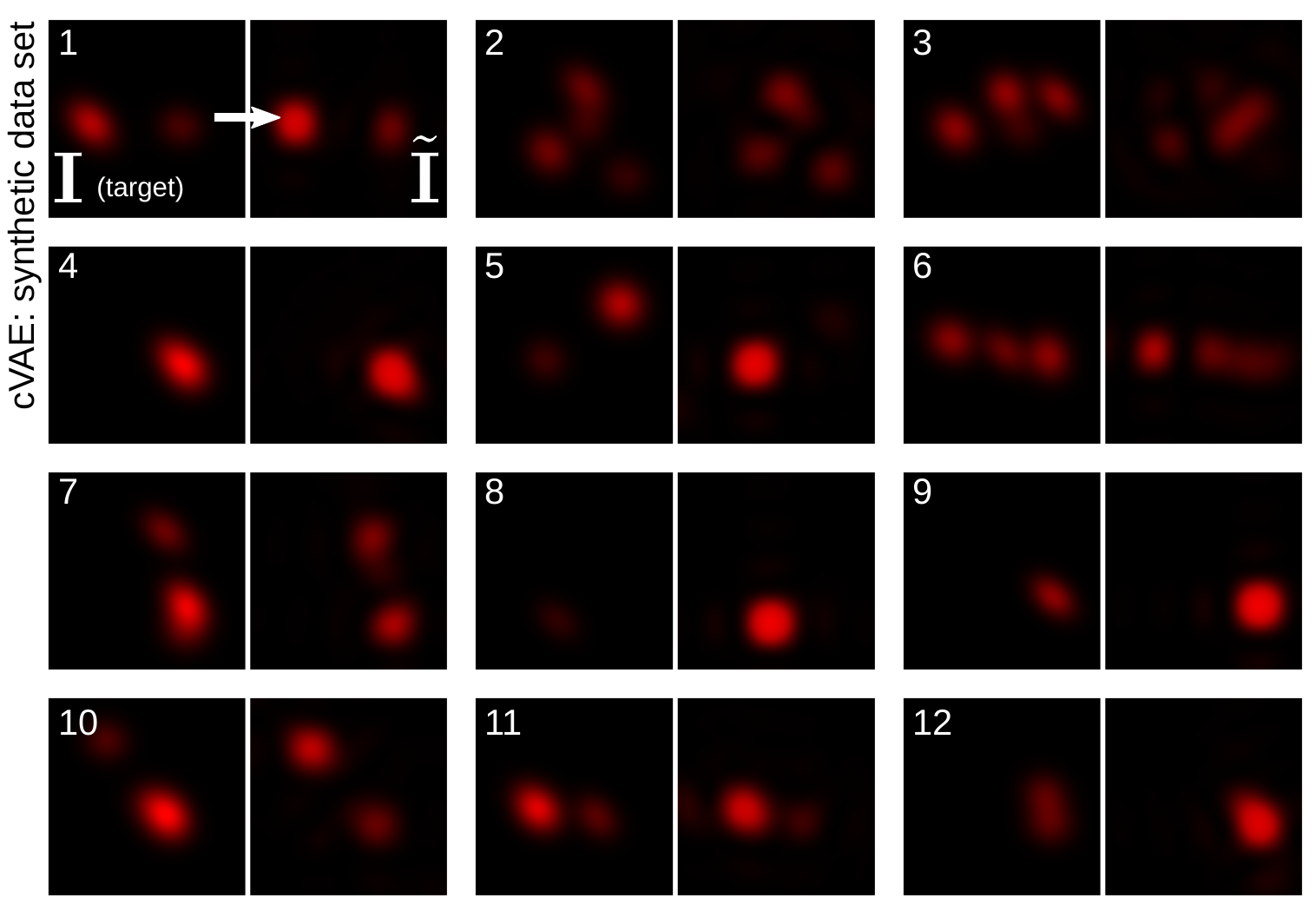}
    \caption{cVAE synthetic data set examples.}
    \label{fig:cVAE_synth_example}
\end{figure}

\begin{figure}
    \centering
    \includegraphics[width=154mm]{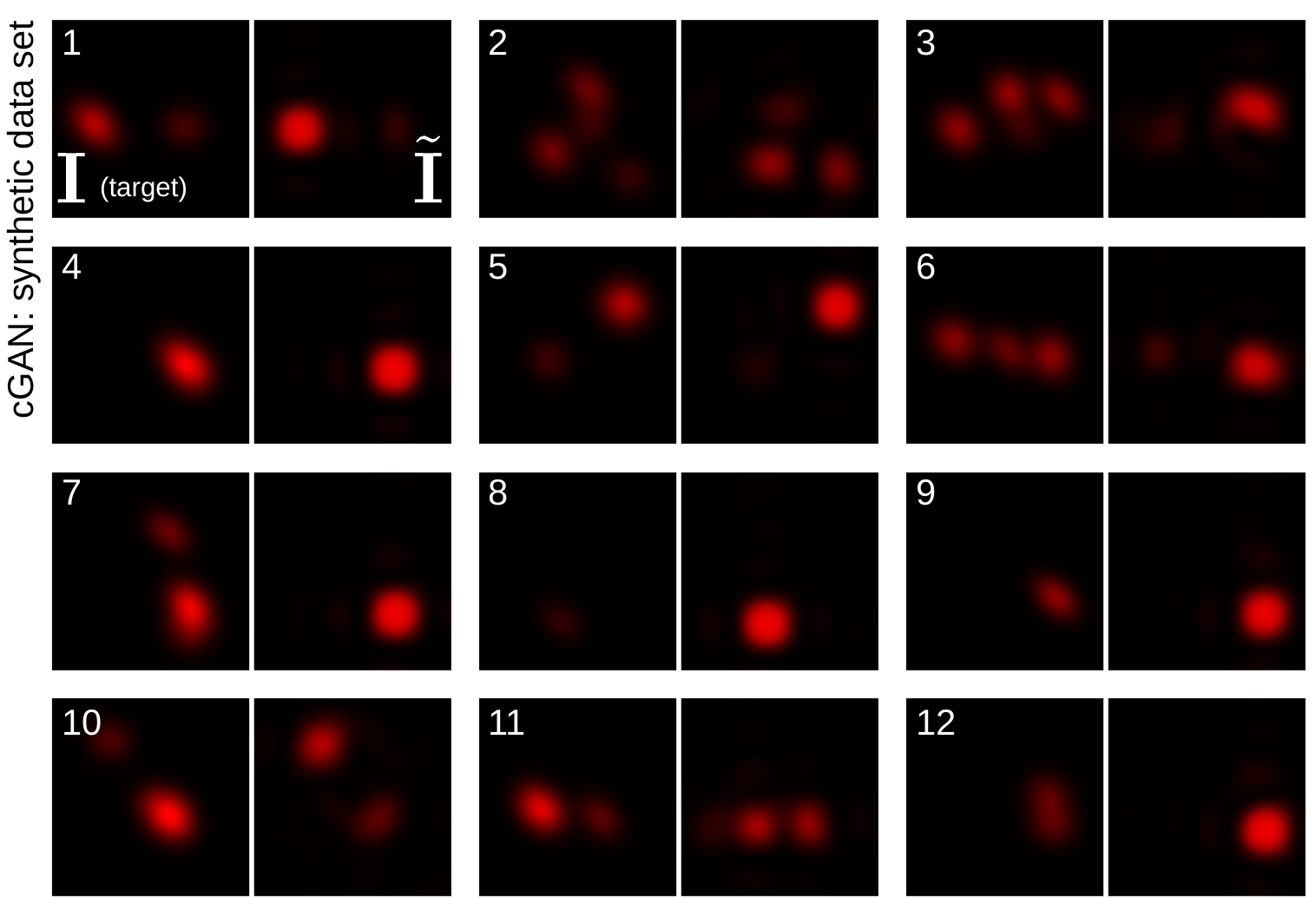}
    \caption{cGAN synthetic data set examples.}
    \label{fig:cGAN_synth_example}
\end{figure}

\begin{figure}
    \centering
    \includegraphics[width=154mm]{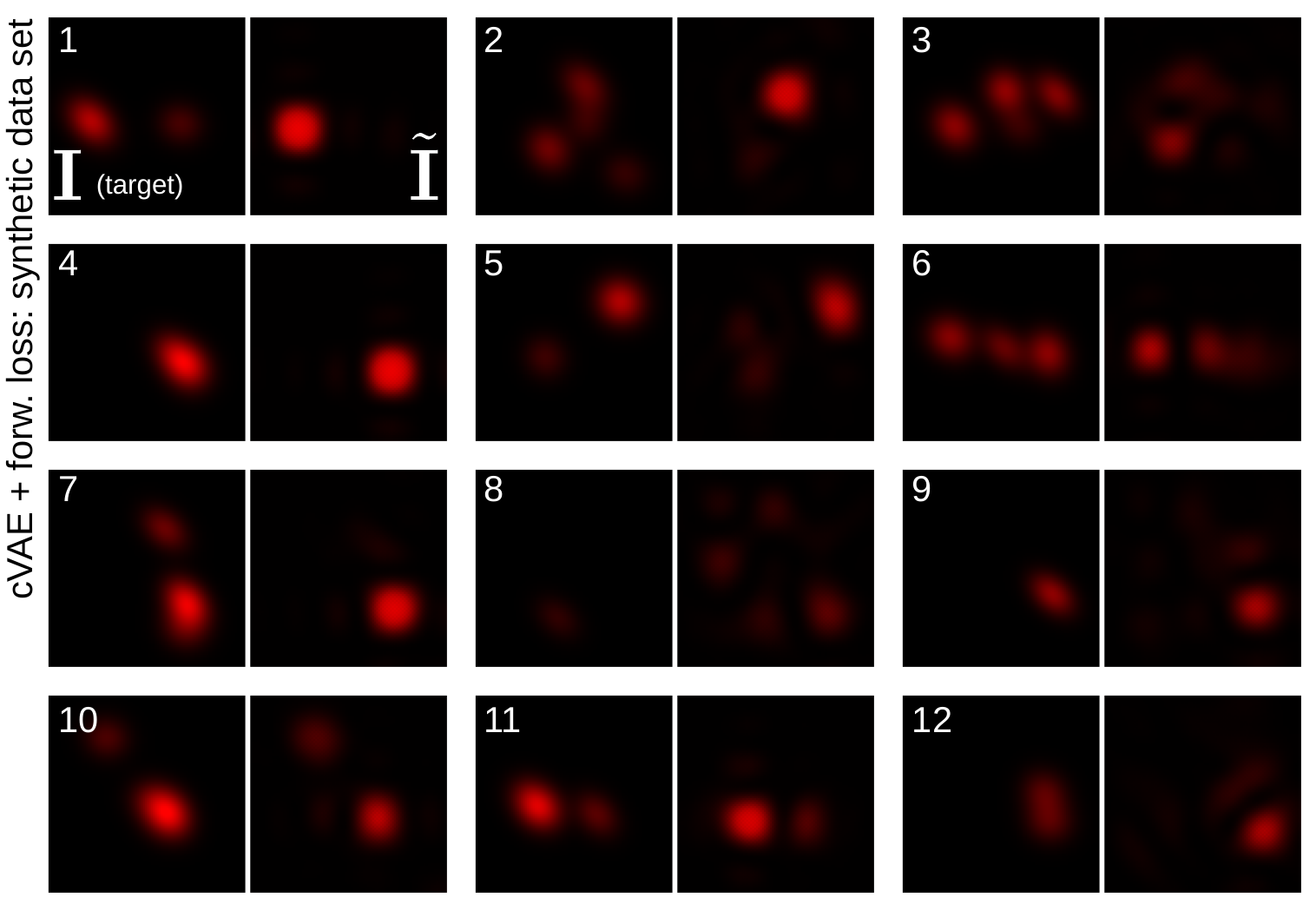}
    \caption{cVAE trained with forward loss synthetic data set examples.}
    \label{fig:cVA_forwE_synth_example}
\end{figure}

\begin{figure}
    \centering
    \includegraphics[width=154mm]{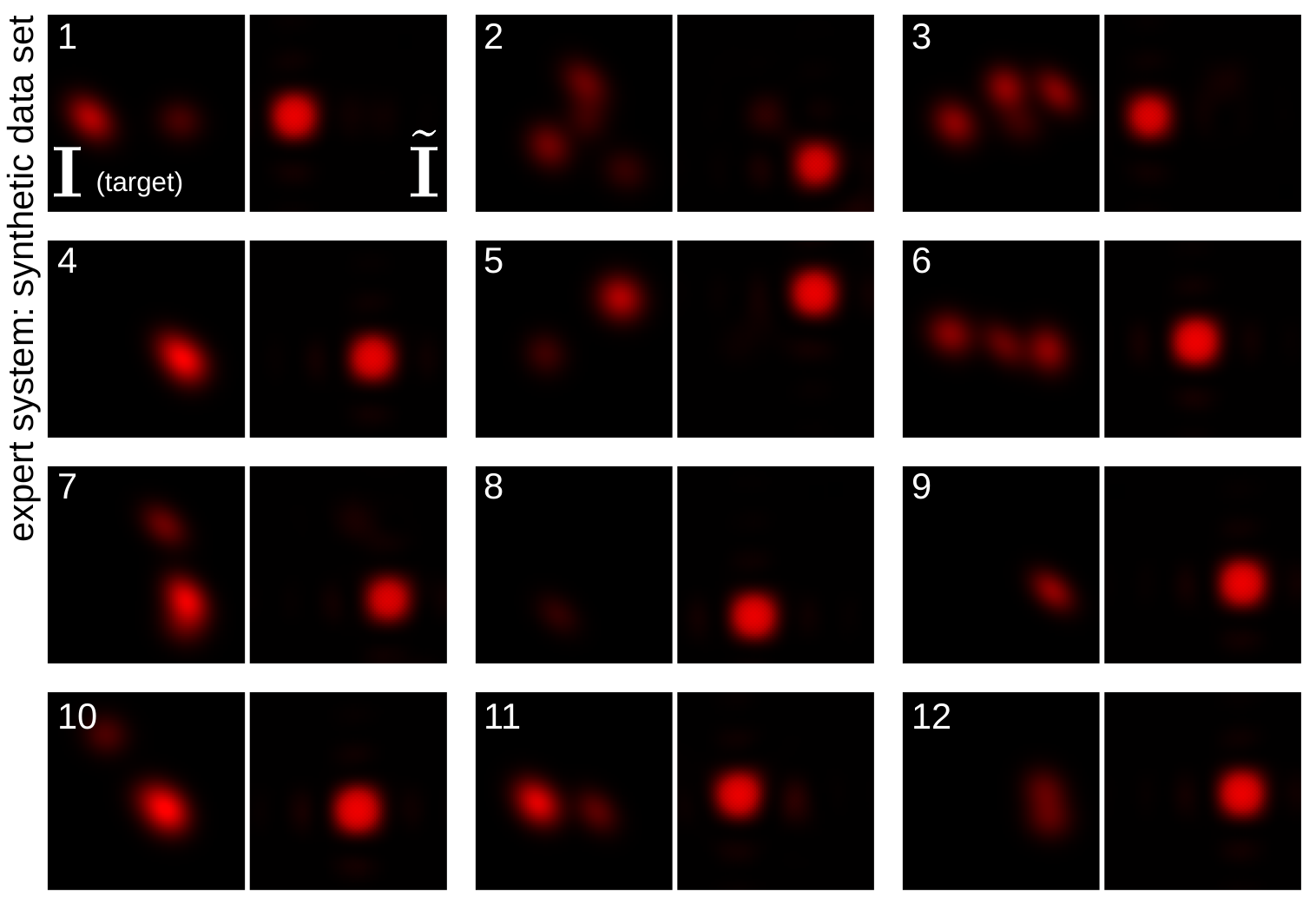}
    \caption{Expert system synthetic data set examples.}
    \label{fig:expert_synth_example}
\end{figure}

In Fig.~\ref{fig:testset_overview} we give an overview of the minimum intensity errors achieved over 5 latent space draws by the three generative models and 5 phase redraws in the case of the expert system. In the left panels, the maximum intensity error is plotted in grey, similar to Fig.~\ref{fig:validation_overview}. As panel {\bf c} shows, the cVAE with forward loss achieves a consistently low intensity error, which is reflected in the height of the peak around $1\%$ in the error distribution. The figure furthermore shows that the difference between the minimum and maximum (grey) intensity error in all four models is low. In order to give a visual impression of model performances we give 12 examples from the synthetic data set and model reconstructions in Figs.~\ref{fig:cVAE_synth_example}-\ref{fig:expert_synth_example}. 

In example 8, the target consists of a single, dim spot. Such a pattern is impossible to create with a single $\mathbf{f}$-matrix element due to the gain-intensity invariance for single non-zero matrix elements that we mentioned in Sec.~\ref{sec:digital_holography} (see also first two examples in Fig.~\ref{fig:fig2}{\bf a}). Indeed, the cVAE, cGAN and the expert system fall into this trap and attempt to simply create a single bright spot at the position of the synthetic peak. The cVAE with forward loss, however, creates multiple dark spots, which results in overall lower error. A similar effect can be observed in example 9.
However, the cVAE with forward loss does not always seem to produce the best fits. In example 2, the standard cVAE manages to produce a convincing approximation of the complex-shaped synthetic intensity distribution. The cVAE with forward loss, by contrast, produces a single spot with a small second maximum. Example 3 appears to be beyond the ability of all four models. 

\newpage
\section{Conclusion}
\label{sec:conclusion}

In this paper, we introduce a, to our knowledge, novel, physics-free approach to the phase-retrieval problem encountered in digital holography based on deep conditional generative models. Our approach is geared towards applications in holographic optical tweezers where intensity distributions often span only a few diffraction-limited spot sizes.

The conditional generative models discussed in this paper are trained on optical data sets prerecorded on our holographic tweezers setup. Importantly, instead of training the models directly on the relation between hologram pixels $\mathbf{h}$ and recorded intensity distributions $\mathbf{I}$, we consider a subset of all possible holograms that can be computed from a complex-valued $8\times 8$ weighting matrix $\mathbf{f}$. 
For each target intensity $\mathbf{I}$, we draw several latent space vectors $z$ at random and consider all resultant $\mathbf{f}$-matrix candidates in terms of the error to the target intensity $E_\mathbf{I}$ that they achieve. 

As we show, the models indeed manage to broadly reproduce intensity patterns, exhibit a promising ability to generalise, and retain some $\mathbf{f}$-matrix variability over the latent space $Z$ as the difference in variability in $E_\mathbf{I}$ and $E_\mathbf{f}$ in Fig.~\ref{fig:validation_overview} indicates. We compare our models with an expert system introduced in Sec.~\ref{sec:model_overview}, which tries to directly infer the $\mathbf{f}$-matrix coefficient corresponding to each local maximum in the intensity distribution using a linear scale. On each data set at least one generative model outperforms the expert system by a margin (see Tbl.~\ref{tab:int_error_overview}).

In our study, we devise cVAE and cGAN models and compare them on the task at hand.
Except for small differences in performance, we find that cVAE and cGAN models behave qualitatively similar. This may be due to the similarity of architecture, as discussed in the models section Sec.~\ref{sec:model_overview}. In both cases, we set the hyperparameter $\beta$ controlling the reconstruction loss in the $\mathbf{f}$-plane to the same value, so that the only differences between the cGAN and cVAE consist in (1) the additional discriminator loss in the case of the cGAN, (2) the distribution of latent space vectors (uniform vs normal), and (3) the encoder loss in the case of the cVAE. It therefore appears that the discriminator rather hinders than helps the cGAN model. 

In addition, we introduce a third architecture: a cVAE that is trained using a forward loss introduced in Secs.~\ref{sec:model_overview} and~\ref{sec:example_noisy_square}.
The forward loss is an attempt to provide the cVAE model with the ability to interpolate in intensity space rather than $\mathbf{f}$-matrix space. As a result, the model does exhibit better performance on the synthetic data set (second column in Tbl.~\ref{tab:int_error_overview}), which, however, appears to come at a cost in the test set as the slightly higher error in the first column shows. This is potentially caused by a lack of capacity of the model to satisfy both the passed-on intensity error and the $\mathbf{f}$-matrix reconstruction error. However, finding holograms to create synthetically-created intensity distributions is more realistic when it comes to actual holographic optical tweezers applications. The cVAE model trained with forward loss might therefore be preferable for such applications, which may require additional research. However, a thorough investigation into the limits and theoretical foundations of our forward-training approach is beyond this practical study.

Our study is a proof-of-principle that generative conditional models can indeed be used for phase-retrieval applications. We do not claim to have found the optimal network architecture for this task as this would require a more in-depth search of possible architectures. In contrast to many deep learning studies, we sampled our own training and test data sets. We are therefore not limited by the size of the data set. The precision and scope of our model depends on the amount of the time users are willing to spend on sampling and training. The quality of intensity (re)construction could thus potentially be further improved by changing the sampling strategy, enlarging the data set and the model. Going forward, one could devise a model which chooses relevant samples on its own and directly interacts with the optical setup. Instead of asking for one-shot estimates of the optimal $\mathbf{f}$-matrix or hologram $\mathbf{h}$, one could adopt a reinforcement-learning strategy. Such an approach would see the network learn an iterative algorithm of repeated cycles of $\mathbf{f}$-matrix suggestion and intensity measurement to arrive at optimal $\mathbf{f}$-matrices or holograms $\mathbf{h}$. Recent developments in generative modelling may constitute interesting alternatives to the type of model studied here. Invertible neural networks, for example, can be trained on the forward and backward mapping of an inverse problem simultaneously and appear therefore suitable for our problem~\cite{ardizzone2018analyzing}.
\paragraph{Acknowledgements} \label{sec:acknowledgements}
The research leading to these
results has received funding from the European Unions Horizon 2020 research and
innovation program under European Training Network (ETN) Grant No. 674979-NANOTRANS. J.G. furthermore acknowledges support of the Winton Programme for the Physics of Sustainability. 
\bibliographystyle{unsrt}  
\bibliography{references}  
\appendix
\section{Supplementary Material}
\label{sec:supplement}

\subsection{Code repository}
\label{sec:code_repo}
All models are implemented in Tensorflow 1.3 using python 3.7.3 and are publicly available on GitHub, \url{https://github.com/JamesGlare/Holo_gen_models}. 

\subsection{Phase invariance for two non-zero $\mathbf{f}$-matrix elements}
\label{sec:phase_invariance_two}
The invariance of the intensity over a phase $\varphi_0$ of an $\mathbf{f}$-matrix element $\mathbf{f}_{i_1,j_1}e^{i\varphi_0}$ with only one other non-zero element $\mathbf{f}_{i_2,j_2}$ follows from 
\begin{align}
    \mathbf{h}_{m,n} &= \textrm{arg}\left [\mathbf{f}_{i_1,j_2} e^{i\vec{k}_{i_1, j_1}\cdot \vec{r}_{m,n}}e^{i\varphi_0} + \mathbf{f}_{i_2,j_2} e^{i\vec{k}_{i_2, j_2}\cdot \vec{r}_{m,n}} \right] \nonumber \\
    &= \left(\frac{\varphi_0}{2}+ \textrm{arg}\left [ \mathbf{f}_{i_1,j_2} e^{i\vec{k}_{i_1, j_1}\cdot \vec{r}_{m,n}}e^{i\varphi_0/2} + \mathbf{f}_{i_2,j_2} e^{i\vec{k}_{i_2, j_2}\cdot \vec{r}_{m,n}}e^{-i\varphi_0/2} \right]\, \right)\textrm{mod} \, 2\pi \nonumber \\ 
    &= \left(\frac{\varphi_0}{2}+ \mathbf{h}'_{m,n}\right) \, \textrm{mod} \,2\pi  \label{eq:two_element_invariance}
\end{align}
where $\mathbf{h}'_{m,n}$ is equal to $\mathbf{h}_{m,n}$ for $\varphi_0=0$, 
such that $\mathbf{h}'_{m,n} $ and $\mathbf{h}_{m,n}$ fulfil the phase invariance described in the Sec.~\ref{sec:digital_holography}. The convention adopted in this paper is that $\textrm{arg}[z]$ maps to $(0, 2\pi)$.

\subsection{Network architectures and training details}
\label{sec:architecture}

\begin{figure}
    \centering
    \includegraphics[width=154mm]{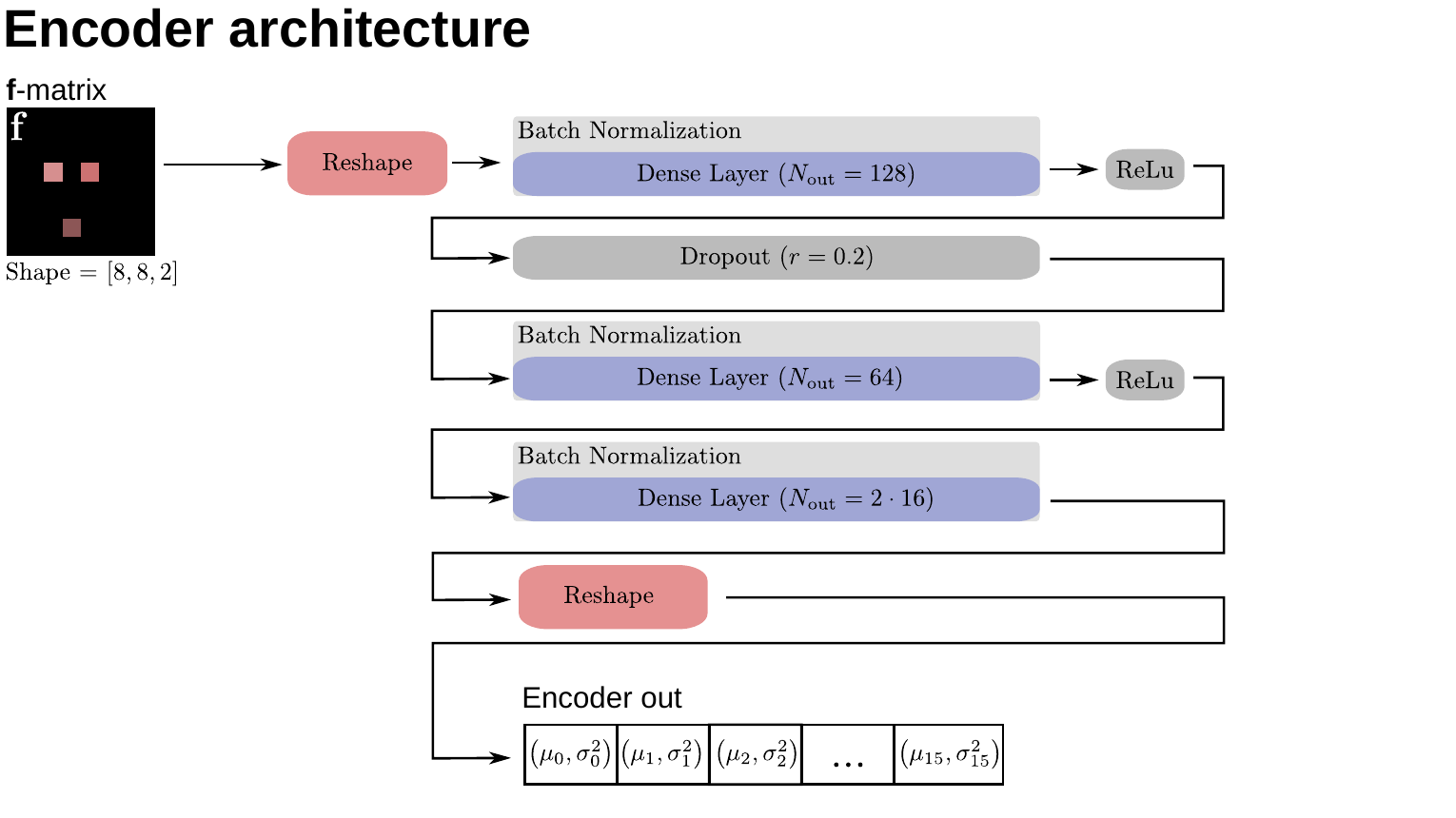}
    \caption{Encoder architecture of the cVAE and cVAE with forward loss.}
    \label{fig:encoder_architecture}
\end{figure}

\begin{figure}
    \centering
    \includegraphics[width=154mm]{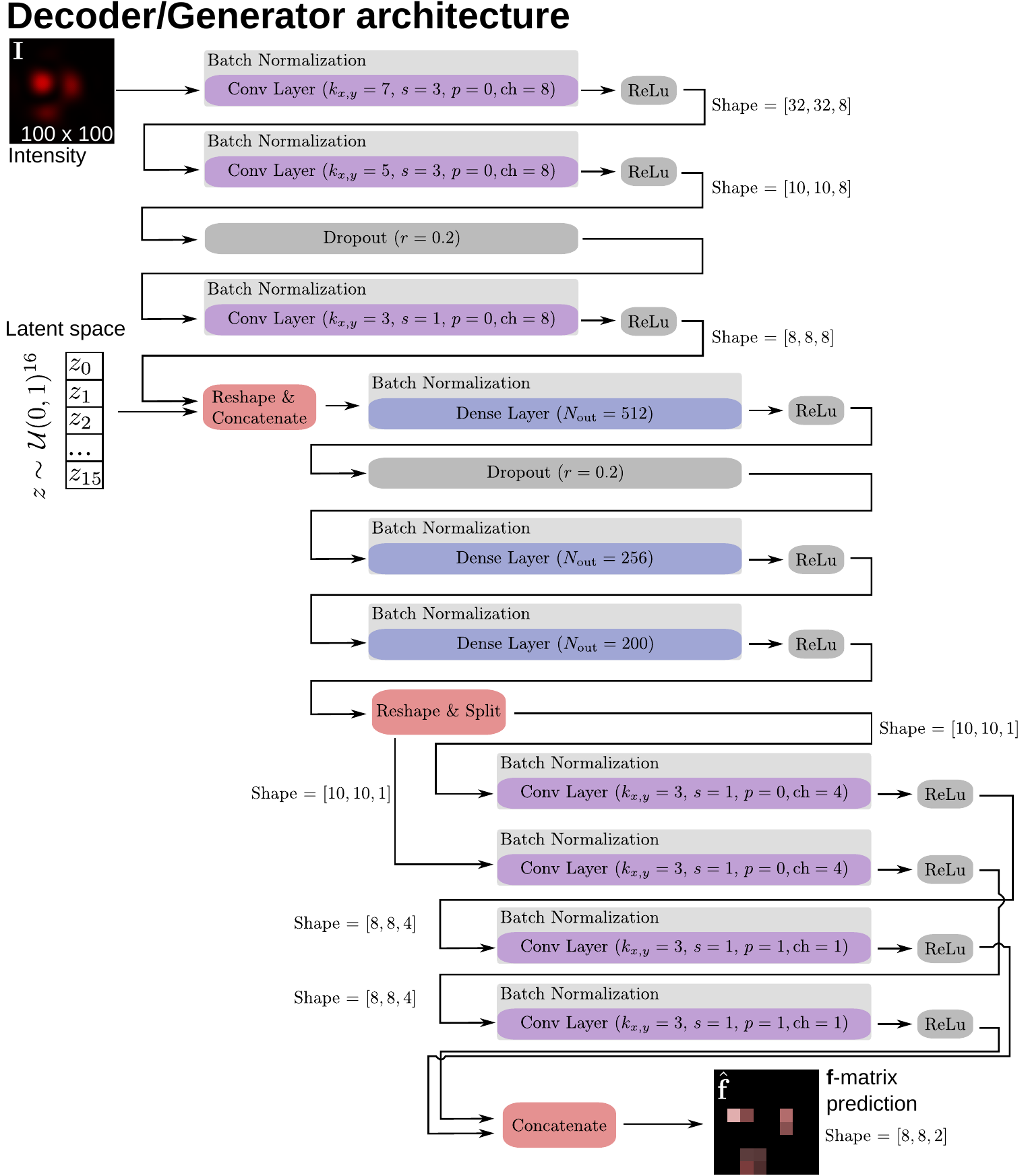}
    \caption{Decoder architecture of the cVAE and cVAE with forward loss. The generator in the cGAN has the exact same architecture.}
    \label{fig:decoder_architecture}
\end{figure}

\begin{figure}
    \centering
    \includegraphics[width=154mm]{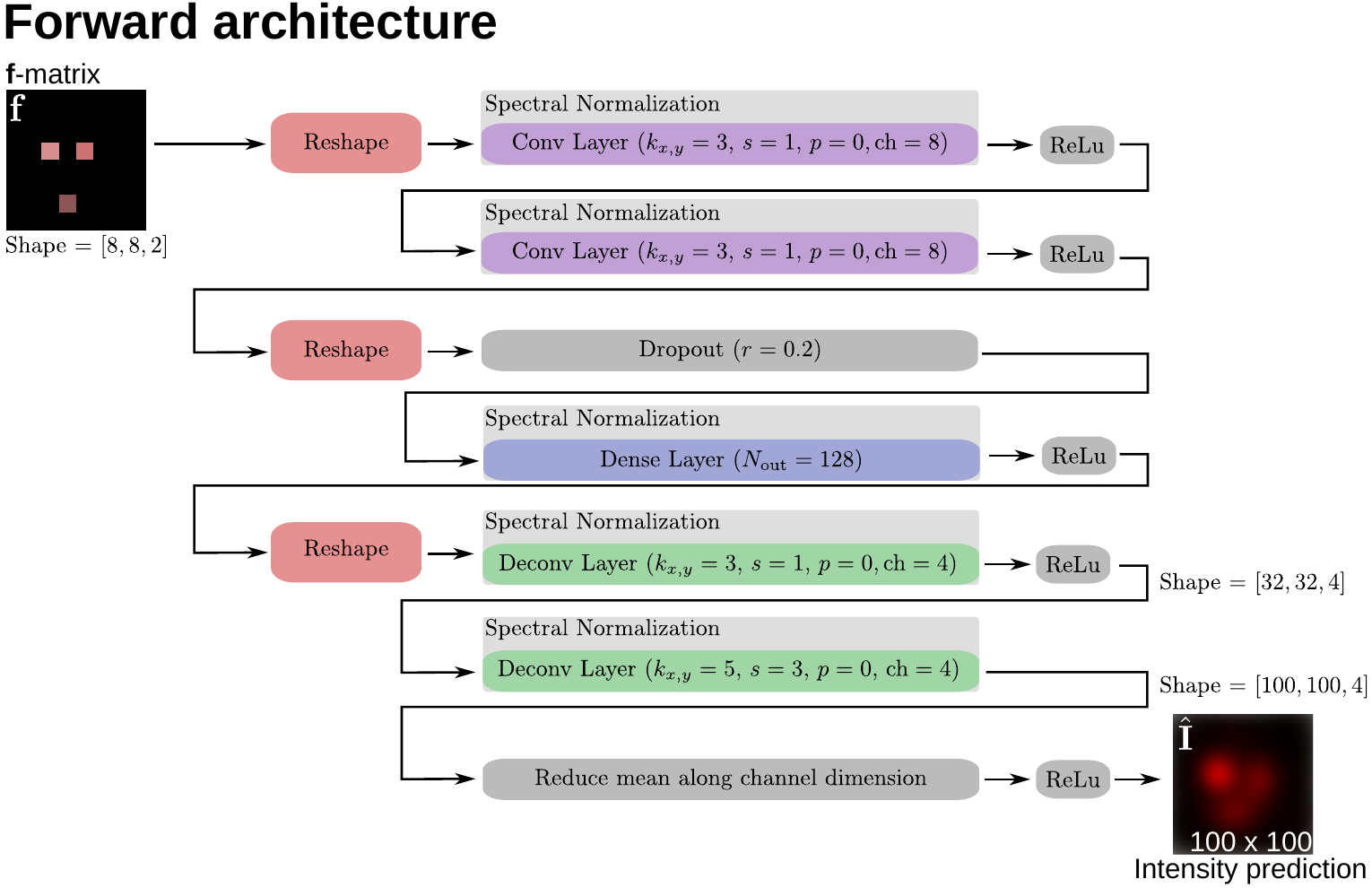}
    \caption{Architecture of the forward network $U_\xi:\mathbf{f}\to\hat{ \mathbf{I}}$.}
    \label{fig:forward_architecture}
\end{figure}

\begin{figure}
    \centering
    \includegraphics[width=154mm]{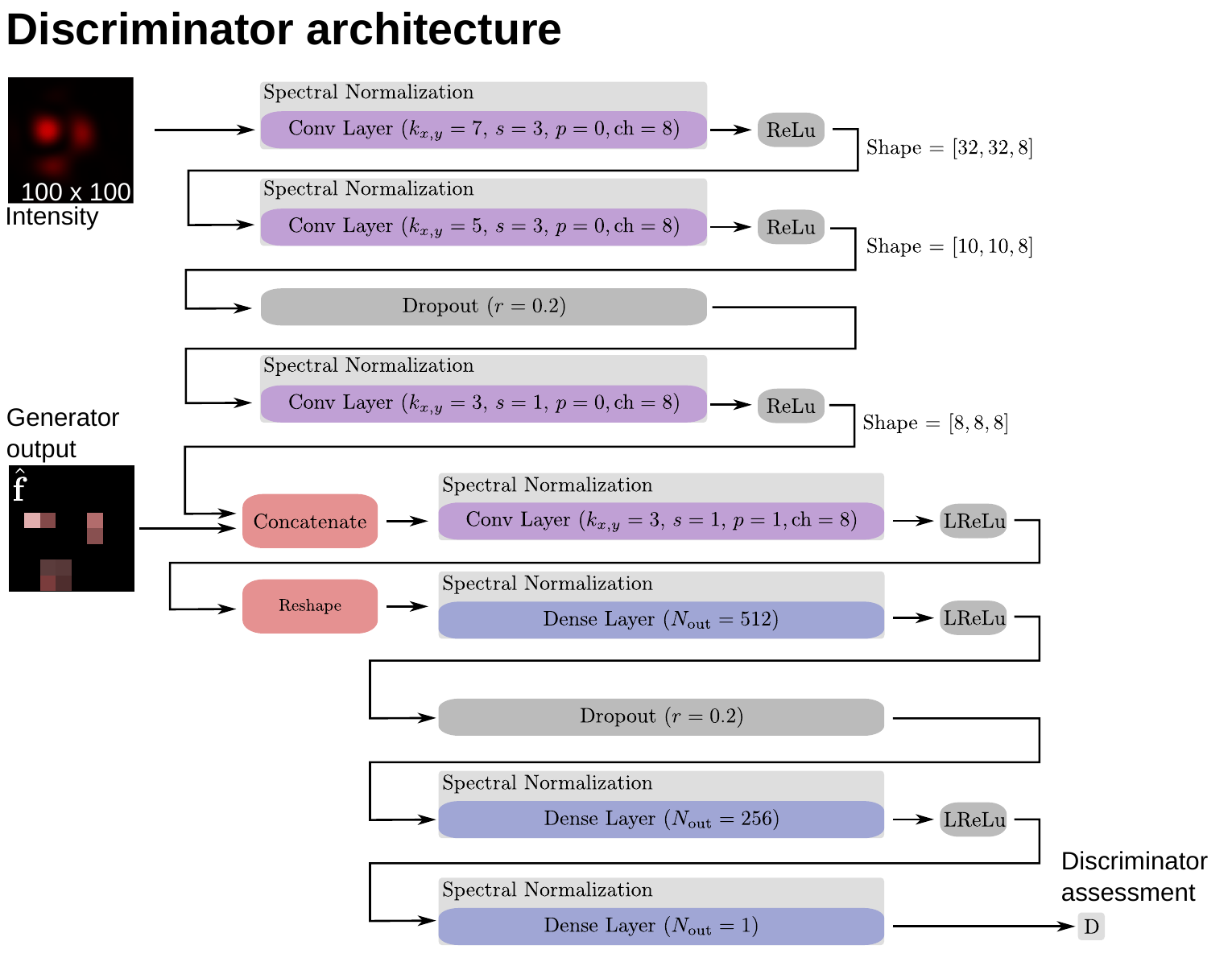}
    \caption{Architecture of the discriminator network used in the cGAN model.}
    \label{fig:discriminator_architecture}
\end{figure}

The network architecture of all models is relatively similar. We give the architecture of the cVAE model in Fig.~\ref{fig:encoder_architecture} (encoder) and Fig.~\ref{fig:decoder_architecture} (decoder). The cVAE with forward loss is exactly similar, with the difference that during training it contains a further network, the forward network shown in Fig.~\ref{fig:forward_architecture}.
The generator in the cGAN architecture is exactly similar to the decoder in Fig.~\ref{fig:decoder_architecture}. The discriminator is given in Fig.~\ref{fig:discriminator_architecture}.

In Tbl.~\ref{tab:training_pars}, we summarise the hyperparameters, choice of optimisers, and training epochs used to train the generative models.
\begin{table}[]
    \centering
    \begin{tabular}{l  c  c  c  }
        Property & cVAE & cGAN & cVAE + forw. loss\\
        \hline
        \hline
        Optimizer: & RMSProp~\cite{tieleman2012lecture}& ADAM~\cite{Kingma2014} & RMSProp/ADAM \\
        Learning rate ($\eta$): & $10^{-4}$ & $10^{-4}$ & $10^{-4}$ \\
        Disc. : Gen. SDG updates: & N.A. & 5 &  1 (gen/forw)\\
        Init. weight mean \& variance: & xavier & xavier & xavier\\
        Latent space dim.& 16 & 16 & 16\\
        Random seed: & 42 & 42 & 42  \\
        Batch size: & 100 & 100 & 100  \\
        Epochs: & 20 & 20 & 20 \\
        \hline
    \end{tabular}
    \caption{Training parameters for all generative models tested in this study.}
    \label{tab:training_pars}
\end{table}


\subsection{Experimental details}
\label{sec:exp_details}
The optical setup used for this study is part of a holographic optical tweezers setup described in~\cite{Gladrow2019}. For the purpose of clarity, we simplified the optical setup in Fig.~\ref{fig:fig1}: Instead of using a single lens placed upstream to the CMOS camera, we branch out a fraction of the laser beam using a 2'' 8:92-pellicle beamsplitter. This fraction of the laser passes three consecutive lenses in 2f-configuration downstream from the SLM, which has a comparable effect to passing only a single lens. The respective focal lengths are $f_1 = 500$ mm, $f_2=200$ mm, and $f_3=100$ mm.

The camera shown in Fig.~\ref{fig:fig1} is a USB-monochrome CMOS with $1280\times 1024$ pixels (Imagingsource). The gain and exposure time in all experiments was set to $8X$ and $10$ ms respectively. The $100\times 100$-sized region-of-interest we choose here as intensity distribution readout corresponds to an $6.2\times 6.2$ $\mu$m$^2$-area in the microscope plane of the holographic tweezers setup.

The laser used here is an Ytterbium fiber laser (\emph{YLM-5- 1064-LP}, IPG photonics) with a wavelength of $\lambda= 1064$~nm, which we operated at an output power of $P=1$~W. We used a $800\times 600$-pixel SLM (LCOS X10468, Hamamatsu) with a refresh rate of 60 Hz.    
\end{document}